\documentstyle[12pt,graphicx,amsmath]{article}
\newlength{\pubnumber} 

\catcode`\@=11
\@addtoreset{equation}{section}

\def\section{\@startsection{section}{1}{\z@}{3.5ex plus 1ex minus .2ex}
 {2.3ex plus .2ex}{\large\bf}}
\def\subsection{\@startsection{subsection}{2}{\z@}{2.3ex plus .2ex}
 {2.3ex plus .2ex}{\bf}}

\newcommand{\oo}[2]{\left(#1\left|#2\right.\right)}
\newcommand{\ba}{\begin{eqnarray}}
\newcommand{\ea}{\end{eqnarray}}
\begin{document}

\begin{titlepage}
\samepage{
\setcounter{page}{1}
\rightline{LPTENS--10/26}
\rightline{LTH--878}
\rightline{July 2010}

\vfill
\begin{center}
 {\Large \bf
Classification of Heterotic Pati--Salam Models
}
\vspace{1cm}
\vfill {\large
Benjamin Assel$^{1}$,
Kyriakos Christodoulides$^{2}$\\
\vspace{.1in}
Alon E. Faraggi$^{2}$,
Costas Kounnas$^{3}$\footnote{Unit\'e Mixte de Recherche
(UMR 8549) du CNRS et de l'ENS
 associ\'eÀe \`a l'universit\'eÀ Pierre et Marie Curie}
 and
John Rizos$^{4}$}\\
\vspace{1cm}
{\it $^{1}$ Centre de Physique Th\'eorique,
             Ecole Polytechnique,
         F--91128 Palaiseau, France\\}
\vspace{.05in}
{\it $^{2}$ Dept.\ of Mathematical Sciences,
             University of Liverpool,
         Liverpool L69 7ZL, UK\\}
\vspace{.05in}
{\it $^{3}$ Lab.\ Physique Th\'eorique,
Ecole Normale Sup\'erieure, F--75231 Paris 05, France\\}
\vspace{.05in}
{\it $^{4}$ Department of Physics,
              University of Ioannina, GR45110 Ioannina, Greece\\}
\vspace{.025in}
\end{center}
\vfill
\begin{abstract}

We extend the classification of free fermionic heterotic--string models
to vacua in which the $SO(10)$ GUT symmetry is broken at the string level
to the Pati--Salam subgroup. Using our classification  method we recently
presented the first example of a quasi--realistic heterotic--string
vacuum that is free of massless exotic states. Within this method
we are able to derive
algebraic expressions for the Generalised GSO (GGSO)
projections for all sectors
that appear in the models. This facilitates
the programming of the entire spectrum analysis in a computer code.
The total number of vacua in the class of models that we
classify is $2^{51}\sim10^{15}$. We perform a statistical sampling
in this space of models and extract $10^{11}$ GGSO configurations
with Pati--Salam gauge group.
Our results demonstrate that one in every $10^{6}$ vacua correspond to a
three generation exophobic model with the required Higgs states,
needed to induce spontaneous breaking to the Standard Model.

\noindent

\end{abstract}
\smallskip}
\end{titlepage}

\setcounter{footnote}{0}

\def\beq{\begin{equation}}
\def\eeq{\end{equation}}
\def\beqn{\begin{eqnarray}}
\def\eeqn{\end{eqnarray}}

\def\no{\noindent }
\def\nolabel{\nonumber }
\def\ie{{\it i.e.}}
\def\eg{{\it e.g.}}
\def\half{{\textstyle{1\over 2}}}
\def\third{{\textstyle {1\over3}}}
\def\quarter{{\textstyle {1\over4}}}
\def\sixth{{\textstyle {1\over6}}}
\def\m{{\tt -}}
\def\p{{\tt +}}

\def\Tr{{\rm Tr}\, }
\def\tr{{\rm tr}\, }

\def\slash#1{#1\hskip-6pt/\hskip6pt}
\def\slk{\slash{k}}
\def\GeV{\,{\rm GeV}}
\def\TeV{\,{\rm TeV}}
\def\y{\,{\rm y}}
\def\SM{Standard--Model }
\def\SUSY{supersymmetry }
\def\SSSM{supersymmetric standard model}
\def\vev#1{\left\langle #1\right\rangle}
\def\l{\langle}
\def\r{\rangle}
\def\o#1{\frac{1}{#1}}

\def\Htw{{\tilde H}}
\def\chibar{{\overline{\chi}}}
\def\qbar{{\overline{q}}}
\def\ibar{{\overline{\imath}}}
\def\jbar{{\overline{\jmath}}}
\def\Hbar{{\overline{H}}}
\def\Qbar{{\overline{Q}}}
\def\abar{{\overline{a}}}
\def\alphabar{{\overline{\alpha}}}
\def\betabar{{\overline{\beta}}}
\def\tautwo{{ \tau_2 }}
\def\thetatwo{{ \vartheta_2 }}
\def\thetathree{{ \vartheta_3 }}
\def\thetafour{{ \vartheta_4 }}
\def\ttwo{{\vartheta_2}}
\def\tthree{{\vartheta_3}}
\def\tfour{{\vartheta_4}}
\def\ti{{\vartheta_i}}
\def\tj{{\vartheta_j}}
\def\tk{{\vartheta_k}}
\def\calF{{\cal F}}
\def\smallmatrix#1#2#3#4{{ {{#1}~{#2}\choose{#3}~{#4}} }}
\def\ab{{\alpha\beta}}
\def\Minv{{ (M^{-1}_\ab)_{ij} }}
\def\bone{{\bf 1}}
\def\ii{{(i)}}
\def\V{{\bf V}}
\def\N{{\bf N}}

\def\b{{\bf b}}
\def\S{{\bf S}}
\def\X{{\bf X}}
\def\I{{\bf I}}
\def\mb{{\mathbf b}}
\def\mS{{\mathbf S}}
\def\mX{{\mathbf X}}
\def\mI{{\mathbf I}}
\def\balpha{{\mathbf \alpha}}
\def\bbeta{{\mathbf \beta}}
\def\bgamma{{\mathbf \gamma}}
\def\bxi{{\mathbf \xi}}

\def\t#1#2{{ \Theta\left\lbrack \matrix{ {#1}\cr {#2}\cr }\right\rbrack }}
\def\C#1#2{{ C\left\lbrack \matrix{ {#1}\cr {#2}\cr }\right\rbrack }}
\def\tp#1#2{{ \Theta'\left\lbrack \matrix{ {#1}\cr {#2}\cr }\right\rbrack }}
\def\tpp#1#2{{ \Theta''\left\lbrack \matrix{ {#1}\cr {#2}\cr }\right\rbrack }}
\def\l{\langle}
\def\r{\rangle}
\newcommand{\cc}[2]{c{#1\atopwithdelims[]#2}}
\newcommand{\nn}{\nonumber}


\def\inbar{\,\vrule height1.5ex width.4pt depth0pt}

\def\IC{\relax\hbox{$\inbar\kern-.3em{\rm C}$}}
\def\IQ{\relax\hbox{$\inbar\kern-.3em{\rm Q}$}}
\def\IR{\relax{\rm I\kern-.18em R}}
 \font\cmss=cmss10 \font\cmsss=cmss10 at 7pt
\def\IZ{\relax\ifmmode\mathchoice
 {\hbox{\cmss Z\kern-.4em Z}}{\hbox{\cmss Z\kern-.4em Z}}
 {\lower.9pt\hbox{\cmsss Z\kern-.4em Z}}
 {\lower1.2pt\hbox{\cmsss Z\kern-.4em Z}}\else{\cmss Z\kern-.4em Z}\fi}

\def\AEF{A.E. Faraggi}
\def\JHEP#1#2#3{{\it JHEP}\/ {\bf #1} (#2) #3}
\def\NPB#1#2#3{{\it Nucl.\ Phys.}\/ {\bf B#1} (#2) #3}
\def\PLB#1#2#3{{\it Phys.\ Lett.}\/ {\bf B#1} (#2) #3}
\def\PRD#1#2#3{{\it Phys.\ Rev.}\/ {\bf D#1} (#2) #3}
\def\PRL#1#2#3{{\it Phys.\ Rev.\ Lett.}\/ {\bf #1} (#2) #3}
\def\PRT#1#2#3{{\it Phys.\ Rep.}\/ {\bf#1} (#2) #3}
\def\MODA#1#2#3{{\it Mod.\ Phys.\ Lett.}\/ {\bf A#1} (#2) #3}
\def\IJMP#1#2#3{{\it Int.\ J.\ Mod.\ Phys.}\/ {\bf A#1} (#2) #3}
\def\nuvc#1#2#3{{\it Nuovo Cimento}\/ {\bf #1A} (#2) #3}
\def\RPP#1#2#3{{\it Rept.\ Prog.\ Phys.}\/ {\bf #1} (#2) #3}
\def\EJP#1#2#3{{\it Eur.\ Phys.\ Jour.}\/ {\bf C#1} (#2) #3}
\def\etal{{\it et al\/}}

\hyphenation{su-per-sym-met-ric non-su-per-sym-met-ric}
\hyphenation{space-time-super-sym-met-ric}
\hyphenation{mod-u-lar mod-u-lar--in-var-i-ant}


\setcounter{footnote}{0}
\section{Introduction}

The heterotic--string models constructed in the free fermionic formulation
\cite{fff}
are among the most realistic
string models constructed to date \cite{fsu5,fny,alr,eu,cfn,lrs}.
These models correspond to $Z_2\times Z_2$ (asymmetric)--orbifold
compactifications, based on ${\cal N}=(2,0)$ super--conformal symmetry
on the world--sheet. The fermionic construction is set
at a special extended symmetry point in the moduli space, and where all
compact dimensions are represented in terms of two dimensional free fermions
propagating on the string world--sheet \cite{z2z21,z2z22}. Marginal
deformations from the
free fermionic point can then be explored by incorporating Thirring
interactions among the world--sheet fermions \cite{thirring}. The free
fermionic construction
provides a set of rules that enables straightforward
extraction of the massless states and interactions, and is therefore
particularly suited to explore the phenomenological properties of
string vacua. The quasi--realistic free fermionic $Z_2\times Z_2$ orbifolds
preserve the $SO(10)$ GUT embedding of the Standard Model spectrum.
The matter states arise from spinorial 16 representations, and the
Higgs states arise from the vectorial 10 representation. It should be
noted that in these models the $SO(10)$ symmetry is broken directly at
the string level, rather than in the effective low energy quantum field
theory. The manifest symmetry in the effective low energy
field theory is therefore a subgroup of $SO(10)$.

Early examples of quasi--realistic free fermionic constructions
were obtained in the late eighties \cite{fsu5,fny,alr,eu}.
Over the past few years tools for the systematic
classification of free fermionic $Z_2\times Z_2$ orbifolds were developed.
In the orbifold language \cite{orbifold} the free fermionic construction
corresponds to symmetric, asymmetric or freely acting orbifolds
\cite{z2z21,z2z22,fknr,cfkr}.
A subclass of them corresponds to symmetric $Z_2\times Z_2$ orbifold
compactifications at enhanced symmetry points in the toroidal moduli
space \cite{z2z21,z2z22}. The chiral matter spectrum arises from
twisted sectors and thus does not depend on the moduli. This
facilitates the complete classification of the topological sectors of the
$Z_2\times Z_2$ symmetric orbifolds. For type II string $N=2$
supersymmetric vacua
the general free fermionic classification techniques were developed in ref.
\cite{gkr}. The method was extended in refs.
\cite{fknr,fkr,fkr2,fkr3,cfkr} for the classification of heterotic $Z_2\times
Z_2$ free
fermionic orbifolds, with unbroken $SO(10)$ and $E_6$ GUT symmetries.
The classification of heterotic $N=1$ (and $N=2$) vacua revealed a symmetry
in the distribution of $Z_2\times Z_2$ (and $Z_2$) string vacua
under exchange of vectorial, and spinorial plus anti--spinorial,
representations of $SO(10)$ \cite{fkr,fkr2,fkr3,cfkr,aft},
akin to mirror symmetry \cite{mirror}.

Our classification methodology entails the expression of the Generalised GSO
(GGSO) projections in terms of generic algebraic equations
for the states that arise in the twisted sectors.
The equations are incorporated in a computer code
that allows scanning a large number of models. In ref. \cite{fknr}
models with $N=1$ space--time supersymmetry
that produce spinorial states from all three distinct twisted
sectors of the $Z_2\times Z_2$ orbifold,
were classified with respect to the number of chiral 16
representations. Such models were dubbed $S^3$ models. This was extended in
ref. \cite{fkr} to models that may produce twisted vectorial 10
representations.
Such models were dubbed $S^2V$, $SV^2$ and $V^3$ models,
corresponding to vacua
in which two, one and none, of the twisted sectors produce
spinorial representations.
The novelty of ref. \cite{fkr} was that a single basis is used to generate
the different classes of models, which substantially simplifies
the classification.
All the different classes of models are generated by choices of the GGSO
projection coefficients. This can be compared with the
method of ref. \cite{nooij}
that uses different basis sets to generate the $S^2V$,
$SV^2$ and $V^3$ type of
models. In ref. \cite{fkr2} the classification
was extended to include vectorial
10 representations in the data output. This enabled the observation of the
spinor--vector duality over the entire space of $N=1$ models.
Ref. \cite{fkr3} demonstrated the existence of spinor--vector duality
in $N=2$ models. Ref. \cite{cfkr} elaborated further on the spinor--vector
duality, in particular in terms of the operational interpretation of
the GGSO free phases, and the breaking of the $N=2$ right--moving
world--sheet supersymmetry.

Absence of adjoint Higgs representations
in heterotic--string models with unbroken $SO(10)$ GUT symmetries
realised as level
one Kac--Moody algebras implies that the models classified in \cite{fknr,fkr}
cannot be spontaneously broken to the Standard Model in the effective field
theory level. Thus, the $SO(10)$ GUT gauge symmetry must be
broken directly at the
string level. In the free fermionic models the GUT gauge symmetry generated by
untwisted vector bosons is $SO(10)$, and can be enhanced to a larger gauge
group by gauge bosons arising from other sectors. Phenomenologically the
most appealing case is that of $SO(10)$ by itself, and therefore it is
reasonable to demand that gauge bosons which enhance the $SO(10)$ symmetry
be projected out by the Generalised GSO (GGSO) projections.
The $SO(10)$ symmetry
must therefore be broken to one of its subgroups.
The cases with $SU(5)\times U(1)$
(flipped $SU(5)$) \cite{fsu5}, $SO(6)\times SO(4)$ (Pati--Salam) \cite{alr},
$SU(3)\times SU(2)\times U(1)^2$ (Standard--like) \cite{fny,eu}
and $SU(3)\times SU(2)^2\times U(1)$ (left--right symmetric) \cite{lrs}
were shown to produce quasi--realistic examples.

The Pati--Salam models obtained via the free fermionic construction of
the heterotic--string
utilise only periodic and anti--periodic boundary conditions,
whereas all the other cases necessarily use
fractional boundary conditions as well. The Pati--Salam case \cite{ps}
therefore represents the simplest extension of the classification program
of \cite{fknr,fkr,fkr2,fkr3,cfkr} to quasi--realistic models.
The Pati--Salam string models contain sectors that preserve
the underlying $SO(10)$ symmetry, as well as sectors that break that
symmetry to the Pati--Salam subgroup.
In general, the $SO(10)$ breaking sectors
may contain massless exotic states that carry fractional electric charge
\cite{ww,schellekens}. The existence of such states is severely constrained
by observations \cite{halyo}.

In ref. \cite{exophobic} our
classification
method was used to demonstrate the existence of quasi--realistic string models
that do not contain massless exotic states, which carry fractional electric
charge. In this paper we extend the classification to
Pati--Salam heterotic string models.
The primary benefit of our method is in the representation of the
GGSO projections in algebraic form for all the twisted sectors that a
priori produce
massless states. We can readily extract the full massless spectrum of
these models. The algebraic formulas are
incorporated in a computer code which enables us to scan a
large space of models.

\section{Pati--Salam Heterotic--String Models}\label{analysis}

The free fermionic formulation of the four
dimensional heterotic string in the
light-cone gauge is described
by $20$ left moving  and $44$ right moving real fermions.
A large number of models can be constructed by choosing
different phases picked up by   fermions ($f_A, A=1,\dots,44$) when transported
along the torus non-contractible loops.
Each model corresponds to a particular choice of fermion phases consistent with
modular invariance
that can be generated by a set of  basis vectors $v_i,i=1,\dots,N$
$$v_i=\left\{\alpha_i(f_1),\alpha_i(f_{2}),\alpha_i(f_{3}))\dots\right\}$$
describing the transformation  properties of each fermion
\begin{equation}
f_A\to -e^{i\pi\alpha_i(f_A)}\ f_A, \ , A=1,\dots,44
\end{equation}
The basis vectors span a space $\Xi$ which consists of $2^N$ sectors that give
rise to the string spectrum. Each sector is given by
\begin{equation}
\xi = \sum N_i v_i,\ \  N_i =0,1
\end{equation}
The spectrum is truncated by a generalized GSO projection whose action on a
string state  $|S>$ is
\begin{equation}\label{eq:gso}
e^{i\pi v_i\cdot F_S} |S> = \delta_{S}\ \cc{S}{v_i} |S>,
\end{equation}
where $F_S$ is the fermion number operator and $\delta_{S}=\pm1$ is the
space--time spin statistics index.
Different sets of projection coefficients $\cc{S}{v_i}=\pm1$ consistent with
modular invariance give
rise to different models. Summarizing: a model can be defined uniquely by a set
of basis vectors $v_i,i=1,\dots,N$
and a set of $2^{N(N-1)/2}$ independent projections coefficients
$\cc{v_i}{v_j}, i>j$.

The free fermions in the light-cone gauge in the usual notation are:
$\psi^\mu, \chi^i,y^i, \omega^i, i=1,\dots,6$ (left-movers) and
$\bar{y}^i,\bar{\omega}^i, i=1,\dots,6$,
$\psi^A, A=1,\dots,5$, $\bar{\eta}^B, B=1,2,3$, $\bar{\phi}^\alpha,
\alpha=1,\ldots,8$ (right-movers).
The class of models we investigate, is
generated by a set of thirteen basis vectors
$$
B=\{v_1,v_2,\dots,v_{13}\},
$$
where
\begin{eqnarray}
v_1=1&=&\{\psi^\mu,\
\chi^{1,\dots,6},y^{1,\dots,6}, \omega^{1,\dots,6}| \nonumber\\
& & ~~~\bar{y}^{1,\dots,6},\bar{\omega}^{1,\dots,6},
\bar{\eta}^{1,2,3},
\bar{\psi}^{1,\dots,5},\bar{\phi}^{1,\dots,8}\},\nonumber\\
v_2=S&=&\{\psi^\mu,\chi^{1,\dots,6}\},\nonumber\\
v_{2+i}=e_i&=&\{y^{i},\omega^{i}|\bar{y}^i,\bar{\omega}^i\}, \
i=1,\dots,6,\nonumber\\
v_{9}=b_1&=&\{\chi^{34},\chi^{56},y^{34},y^{56}|\bar{y}^{34},
\bar{y}^{56},\bar{\eta}^1,\bar{\psi}^{1,\dots,5}\},\label{basis}\\
v_{10}=b_2&=&\{\chi^{12},\chi^{56},y^{12},y^{56}|\bar{y}^{12},
\bar{y}^{56},\bar{\eta}^2,\bar{\psi}^{1,\dots,5}\},\nonumber\\
v_{11}=z_1&=&\{\bar{\phi}^{1,\dots,4}\},\nonumber\\
v_{12}=z_2&=&\{\bar{\phi}^{5,\dots,8}\},\nonumber\\
v_{13}=\alpha &=& \{\bar{\psi}^{4,5},\bar{\phi}^{1,2}\}.
\nonumber
\end{eqnarray}
The first twelve vectors in this set are identical to
those used in \cite{fknr,fkr}.
The vectors $1,S$ generate an
$N=4$ supersymmetric model, with $SO(44)$ gauge symmetry.
The vectors $e_i,i=1,\dots,6$ give rise
to all possible symmetric shifts of the six internal fermionized coordinates
($\partial X^i=y^i\omega^i, {\bar\partial} X^i= \bar{y}^i\bar{\omega}^i$).
Their addition breaks the $SO(44)$ gauge group, but preserves
$N=4$ supersymmetry.
The vectors $b_1$ and $b_2$  define the $SO(10)$ gauge symmetry and
the $Z_2\times Z_2$ orbifold twists, which break
$N=4$ to $N=1$ supersymmetry.
The $z_1$ and $z_2$ basis vectors reduce the untwisted gauge group generators
from $SO(16)$ to $SO(8)_1\times SO(8)_2$.
Finally $v_{13}$ is the additional new vector that breaks the $SO(10)$ GUT
symmetry to
$SO(6)\times SO(4)$, and the $SO(8)_1$ hidden symmetry to
$SO(4)_1\times SO(4)_2$.

The second ingredient that is needed to
define the string vacuum are the GGSO projection coefficients that
appear in the one--loop partition function,
$\cc{v_i}{v_j}$, spanning a $13\times 13$ matrix.
Only the elements with $i>j$ are
independent while the others are fixed by modular invariance.
A priori there are therefore 78 independent coefficients corresponding
to $2^{78}$ string vacua. Eleven coefficients
are fixed by requiring that the models possess $N=1$ supersymmetry.
Without loss of generality we impose the associated GGSO projection
coefficients
\ba
\cc{1}{1} = \cc{S}{1} = \cc{S}{e_{i}} = \cc{S}{b_{m}} =
\cc{S}{z_n} = \cc{S}{\alpha} = -1 , &&\\
i=1,...,6, \, m = 1,2 , \, n = 1,2. &&\nn
\ea
leaving 66 independent coefficients,
\begin{eqnarray}
&&\cc{e_i}{e_j}, i\ge j, \ \ \cc{b_1}{b_2}, \ \ \cc{z_1}{z_2}, \ \
\cc{1}{b_A}, \ \ \cc{1}{z_A}\nn\\
&&\cc{e_i}{z_n}, \cc{e_i}{b_A},\cc{b_A}{z_n},
\ \  \cc{1}{\alpha}, \cc{e_i}{\alpha}, \cc{b_A}{\alpha}, \cc{z_A}{\alpha}
\nn\\
i,j=1,\dots6\,\ ,\  A,B,m,n=1,2\nn,
\end{eqnarray}
since all of the remaining projection coefficients are determined by modular
invariance \cite{fff}.
Each of the 66 independent coefficients can take two discrete
values $\pm1$ and thus a simple counting gives $2^{66}$
(that is approximately $10^{19.9}$) models in the
class of superstring vacua under consideration.
We remark here that there may exist some degeneracies in this space
of physical vacua with respect to the properties of the effective low 
energy field theory, {\it i.e.} in particular with respect to the
massless spectra. For example, there exists a cyclic permutation symmetry
among the three twisted sectors of the $Z_2\times Z_2$ orbifold.
However, many of the vacua that may seem equivalent from the point
of view of the effective field theory limit of the observable massless spectra,
may differ by other properties, like, for example:
hidden sector matter states; the massive spectrum; superpotential couplings; 
and are therefore  distinct. 
%
%
The important question that we address by a statistical analysis in this paper
is the frequency by which exophobic vacua occur in the total space of 
configurations.

The vector bosons from the untwisted sector generate an
$$
SO(6)\times SO(4)\times{U(1)}^3\times{SO(4)}^2\times SO(8)
$$
gauge symmetry.
Depending on the  choices of the projection coefficients,
extra gauge bosons may arise from the following ten sectors:
\begin{equation}
\mathbf{G} =
\left\{ \begin{array}{cccccc}
z_1          ,&
z_2          ,&
\alpha       ,&
\alpha + z_1 ,&
              &
                \cr
x            ,&
z_1 + z_2    ,&
\alpha + z_2 ,&
\alpha + z_1 + z_2,&
\alpha + x ,&
\alpha + x + z_1
\end{array} \right\} \label{ggsectors}
\end{equation}
where
\begin{equation}
x = 1 + S + \sum_{i=1}^{6} e_i + z_1 + z_2 =
    \{\bar{\eta}^{123}, \bar{\psi}^{12345}\}.
\label{xvector}
\end{equation}

Vector bosons that arise from these sectors enhance the untwisted
gauge symmetry.
We impose the condition that the only space--time vector bosons
that remain in the spectrum are those
that arise from the untwisted sector.
This restricts further the number of phases,
leaving a total of 51 independent GGSO phases.
The gauge group in these models is therefore:
\beqn
{\rm observable} ~: &~~~~~~~~SO(6)\times SO(4) \times U(1)^3 \nonumber\\
{\rm hidden}     ~: &~~SO(4)^2\times SO(8)~~~~             \nonumber
\eeqn
where the hidden $SO(4)^2\sim  SO(4)_1\times SO(4)_2\sim 
SU(2)_1\times  SU(2)_2\times  SU(2)_3\times  SU(2)_4$.

The untwisted matter is common in these models and is composed of three pairs
of vectorial representations of the observable $SO(6)$ symmetry, and 12 states
that are singlets under the non--Abelian gauge groups.
The chiral matter spectrum arises from the twisted sectors.
The chiral spinorial representations of the observable $SO(6) \times SO(4)$
arise from the sectors:
\begin{eqnarray} \label{eqn:obspin}
B_{pqrs}^{(1)}&=& S + b_1 + p e_3+ q e_4 + r e_5 + s e_6 \nonumber\\
&=&\{\psi^\mu,\chi^{12},(1-p)y^{3}\bar{y}^3,p\omega^{3}\bar{\omega}^3,
(1-q)y^{4}\bar{y}^4,q\omega^{4}\bar{\omega}^4, \nonumber\\
& & ~~~(1-r)y^{5}\bar{y}^5,r\omega^{5}\bar{\omega}^5,
(1-s)y^{6}\bar{y}^6,s\omega^{6}\bar{\omega}^6,\bar{\eta}^1,\bar{\psi}^{1..5}\}
\\
B_{pqrs}^{(2)}&=& S + b_2 + p e_1+ q e_2 + r e_5 + s e_6
\label{chiralspinorials2}
\nonumber\\
B_{pqrs}^{(3)}&=& S + b_3 + p e_1+ q e_2 + r e_3 + s e_4 \nonumber
\end{eqnarray}
where $p,q,r,s=0,1$; $b_3=b_1+b_2+x=1+S+b_1+b_2+\sum_{i=1}^6 e_i+\sum_{n=1}^2
z_n$ and
$x$ is given in eq. (\ref{xvector}).
These sectors give rise to \textbf{16} and $\overline{\textbf{16}}$
representations of $SO(10)$ decomposed under
$SO(6)\times SO(4)\equiv SU(4)\times SU(2)_L\times SU(2)_R$
\beqn
\textbf{16} = & \textbf{(4,~2,~1)} + \textbf{(}\bar{\textbf{4}}\textbf{, 1, 2)}
\nonumber\\
\overline{\textbf{16}} = &\textbf{(}\bar{\textbf{4}}\textbf{, 2, 1)} +
\textbf{(4,~1,~2)} \nonumber
\eeqn

The following sectors give rise to states that transform as representations of
the hidden
gauge group, and are singlets under the observable $SO(10)$ GUT symmetry. These
states
are therefore hidden matter states that arise in the string model, but are not
exotic with respect to electric charge.
The following 48 sectors produce the representations
$\textbf{(({2},{1}),({2},{1}))}$ of
$SU(2)^4 = SO(4)_1 \times SO(4)_2$:
\begin{eqnarray}\label{eqn:hidspin1}
B_{pqrs}^{(4)} &=& B_{pqrs}^{(1)} + x + z_1 \,\,=\,\,
           S + b_{1}+ p e_3+ q e_4 + r e_5 + s e_6 + x + z_1 \nonumber\\
&=&\{\psi^\mu,\chi^{12},(1-p)y^{3}\bar{y}^3,p\omega^{3}\bar{\omega}^3,
            (1-q)y^{4}\bar{y}^4,q\omega^{4}\bar{\omega}^4, \nonumber\\
& & ~~~(1-r)y^{5}\bar{y}^5,r\omega^{5}\bar{\omega}^5,(1-s)y^{6}\bar{y}^6,
        s\omega^{6}\bar{\omega}^6,\bar{\eta}^{23},\bar{\phi}^{1..4}\} \\
B_{pqrs}^{(5)} &=& B_{pqrs}^{(2)} + x + z_1 \,\,=\,\,
         S + b_2 + p e_1+ q e_2 + r e_5 + s e_6 + x + z_1\nonumber\\
B_{pqrs}^{(6)} &=& B_{pqrs}^{(3)} + x + z_1 \,\,=\,\,
                S + b_3 + p e_1+ q e_2 + r e_3 + s e_4 + x + z_1\nonumber
\end{eqnarray}
There are 48 sectors producing spinorial $\textbf{{8}}$
and anti--spinorial $\bar{\textbf{8}}$ representations of the hidden
$SO(8)$ gauge group:
\begin{eqnarray}\label{eqn:hidspin2}
B_{pqrs}^{(7)} &=& B_{pqrs}^{(1)} + x + z_2 \,\,=\,\,
S + b_{1}+ p e_3+ q e_4 + r e_5 + s e_6 + x + z_2 \nonumber\\
&=&\{\psi^\mu,\chi^{12},(1-p)y^{3}\bar{y}^3,p\omega^{3}\bar{\omega}^3,
(1-q)y^{4}\bar{y}^4,q\omega^{4}\bar{\omega}^4, \nonumber\\
& & ~~~(1-r)y^{5}\bar{y}^5,r\omega^{5}\bar{\omega}^5,
(1-s)y^{6}\bar{y}^6,s\omega^{6}\bar{\omega}^6,\bar{\eta}^{23},\bar{\phi}^{5..8}\} \\
B_{pqrs}^{(8)} &=& B_{pqrs}^{(2)} + x + z_2 \,\,=\,\,
S + b_2 + p e_1+ q e_2 + r e_5 + s e_6 + x + z_2 \nonumber\\
B_{pqrs}^{(9)} &=& B_{pqrs}^{(3)} + x + z_2 \,\,=\,\,
S + b_3 + p e_1+ q e_2 + r e_3 + s e_4 + x + z_2 \nonumber
\end{eqnarray}

We note that in these models there are three $SO(4)$
group factors, related with a cyclic symmetry.
 We could have therefore defined one of the other two
$SO(4)$ group as the
observable one, and the other two as the hidden ones.
We follow here the convention
that keeps the group generated by the world--sheet fermions
${\bar\psi}^{4,5}$ as
the observable $SO(4)$ and the ones generated by
${\bar\phi}^{1,2}$ and ${\bar\phi}^{3,4}$
as hidden. The models then give rise to a multitude of sectors that
produce exotic states with fractional electric charge, given by:
\beq
Q_{em} = {1\over\sqrt{6}}T_{15}+{1\over2}I_{3_L}+{1\over2}I_{3_R}
\eeq
where $T_{15}$ is the diagonal generator of $SU(4)/SU(3)$ and
$I_{3_L}$, $I_{3_R}$
are the diagonal generators of $SU(2)_L$, $SU(2)_R$, respectively.
The models then contain the exotic states in the representations:
\begin{align}
({\textbf{4}},{\textbf{1}},{\textbf{1}})+({\bar{\textbf{4}}},
{\textbf{1}},{\textbf{1}}):&\pm\frac{1}{6} \ \text{exotic coloured particles
and singlets}\nn\\
({\textbf{1}},{\textbf{2}},{\textbf{1}}):& \pm\frac{1}{2}\ \text{leptons}\nn\\
 ({\textbf{1}},{\textbf{1}},{\textbf{2}}):& \pm\frac{1}{2}\  \text{singlets}\nn
\end{align}

We now enumerate the sectors that give rise to exotic states.
The states corresponding to the representations
$({\textbf{4}},{\textbf{2}},{\textbf{1}})$,
$({\textbf{4}},{\textbf{1}},{\textbf{2}})$,
$(\bar{{\textbf{4}}},{\textbf{2}},{\textbf{1}})$,
$(\bar{{\textbf{4}}},{\textbf{1}},{\textbf{2}})$ where
${\textbf{4}}$ and $\bar{{\textbf{4}}}$ are spinorial (anti--spinorial)
representations of the observable SO(6), and the $\textbf{2}$ are
doublet representations
of the hidden $SU(2)\times SU(2) = SO(4)_1$, arise from the following sectors:
\begin{eqnarray}\label{eqn:exospin1}
B_{pqrs}^{(10)}&=& B_{pqrs}^{(1)} + \alpha \,\,=\,\,
S + b_1 + p e_3+ q e_4 + r e_5 + s e_6 + \alpha  \nonumber\\
&=&\{\psi^\mu,\chi^{12},(1-p)y^{3}\bar{y}^3,p\omega^{3}\bar{\omega}^3,
(1-q)y^{4}\bar{y}^4,q\omega^{4}\bar{\omega}^4, \nonumber\\
& & ~~~(1-r)y^{5}\bar{y}^5,r\omega^{5}\bar{\omega}^5,(1-s)
y^{6}\bar{y}^6,s\omega^{6}\bar{\omega}^6,\bar{\eta}^1,
\bar{\psi}^{1..3},\bar{\phi}^{1..2}\} \\
B_{pqrs}^{(11)}&=&  B_{pqrs}^{(2)} + \alpha \,\,=\,\,
S + b_2 + p e_1+ q e_2 + r e_5 + s e_6 + \alpha \nonumber\\
B_{pqrs}^{(12)}&=&  B_{pqrs}^{(3)} + \alpha \,\,=\,\,
S + b_3 + p e_1+ q e_2 + r e_3 + s e_4 + \alpha \nonumber
\end{eqnarray}
Similar states $B_{pqrs}^{(13,14,15)}$ arise from the sectors
$B_{pqrs}^{(10,11,12)}+ z_1$
and they correspond to the representations
$({\textbf{4}},{\textbf{2}},{\textbf{1}})$,
$({\textbf{4}},{\textbf{1}},{\textbf{2}})$,
$(\bar{{\textbf{4}}},{\textbf{2}},{\textbf{1}})$,
$(\bar{{\textbf{4}}},{\textbf{1}},{\textbf{2}})$ of
$SO(6)_{obs} \times SO(4)_2$.

The states corresponding to the representations
$(({\textbf{2}},{\textbf{1}}),({\textbf{2}},{\textbf{1}}))$,
$(({\textbf{2}},{\textbf{1}}),({\textbf{1}},{\textbf{2}}))$,
$(({\textbf{1}},{\textbf{2}}),({\textbf{1}},{\textbf{2}}))$ and
$(({\textbf{1}},{\textbf{2}}),({\textbf{2}},{\textbf{1}}))$ transforming under
$SU(2)_L\times SU(2)_R \times SO(4)_1$ arise from the sectors:
\begin{eqnarray}\label{eqn:exospin2}
B_{pqrs}^{(16)}&=& B_{pqrs}^{(1)} + \alpha + x \,\,=\,\,
S + b_1 + p e_3+ q e_4 + r e_5 + s e_6 + \alpha + x \nonumber\\
&=&\{\psi^\mu,x^{12},(1-p)y^{3}\bar{y}^3,
p\omega^{3}\bar{\omega}^3,(1-q)y^{4}\bar{y}^4,q\omega^{4}\bar{\omega}^4,
\nonumber\\
& & ~~~(1-r)y^{5}\bar{y}^5,r\omega^{5}\bar{\omega}^5,
(1-s)y^{6}\bar{y}^6,s\omega^{6}\bar{\omega}^6,\bar{\eta}^2,
\bar{\eta}^3,\bar{\psi}^{4..5},\bar{\phi}^{1..2}\}\\
B_{pqrs}^{(17)}&=&  B_{pqrs}^{(2)} + \alpha + x \,\,=\,\,
S + b_2 + p e_1+ q e_2 + r e_5 + s e_6 + \alpha + x \nonumber\\
B_{pqrs}^{(18)}&=&  B_{pqrs}^{(3)} + \alpha + x \,\,=\,\,
S + b_3 + p e_1+ q e_2 + r e_3 + s e_4 + \alpha + x \nonumber
\end{eqnarray}
Similar states $B_{pqrs}^{(19,20,21)}$ arise from the sectors
$B_{pqrs}^{(16,17,18)}+ z_1$ and they produce analogous  representations under
$SU(2)_L\times SU(2)_R \times SO(4)_2$.

Finally states that transform in vectorial representations are obtained from
sectors that
contain four periodic world--sheet right--moving complex fermions. Massless
states
are obtained in such sectors by acting on the vacuum with a Neveu--Schwarz
right--moving fermionic oscillator. Vectorial representations arise from the
sectors:
\begin{eqnarray}\label{eqn:vecto}
B_{pqrs}^{(1)} + x &=& S + b_1 + p e_3+ q e_4 + r e_5 + s e_6 + x \nonumber\\
&=&\{\psi^\mu,\chi^{12},(1-p)y^{3}\bar{y}^3,p\omega^{3}\bar{\omega}^3,
(1-q)y^{4}\bar{y}^4,q\omega^{4}\bar{\omega}^4, \nonumber\\
& & ~~~(1-r)y^{5}\bar{y}^5,r\omega^{5}\bar{\omega}^5,(1-s)y^{6}
\bar{y}^6,s\omega^{6}\bar{\omega}^6,\bar{\eta}^2, \bar{\eta}^3 \} \\
B_{pqrs}^{(2)} + x &=& S + b_2 + p e_1+ q e_2 + r e_5 + s e_6 + x
               \label{vectorials} \nonumber\\
B_{pqrs}^{(3)} + x &=& S + b_3 + p e_1+ q e_2 + r e_3 + s e_4 + x \nonumber
\end{eqnarray}
and produce the following representations:

\begin{itemize}
\item $\{\bar{\psi}^{123}\}|R>_{pqrs}^{(i)}$, $i = 1,2,3$, where
$|R>_{pqrs}^{(i)}$ is the degenerated Ramond vacuum of the
$B_{pqrs}^{(i)}$ sector.
These states transform as a vectorial representation of SO(6).
\item $\{\bar{\psi}^{45}\}|R>_{pqrs}^{(i)}$, $i = 1,2,3$, where
$|R>_{pqrs}^{(i)}$ is the degenerated Ramond vacuum of the
$B_{pqrs}^{(i)}$ sector.
These states transform as a vectorial representation of SO(4).
\item $\{\bar{\phi}^{12}\}|R>_{pqrs}^{(i)}$, $i = 1,2,3$.
These states transform as a vectorial representation of SO(4).
\item $\{\bar{\phi}^{34}\}|R>_{pqrs}^{(i)}$, $i = 1,2,3$.
These states transform as a vectorial representation of SO(4).
\item $\{\bar{\phi}^{5..8}\}|R>_{pqrs}^{(i)}$, $i = 1,2,3$.
These states transform as a vectorial representation of SO(8).
\item the remaining states in those sectors transform as singlets
of the non--Abelian group factors.
\end{itemize}

It is important to note that the states arising from the sectors
in eq. (\ref{eqn:vecto})
are standard states from the point of view of the Standard Model
charge assignments and grand unification embeddings. The term ``exotic states''
applies only to states that arise due to the ``Wilson line''
breaking of the non--Abelian GUT symmetries in string theory.
In the Pati--Salam models these are the states that arise from the 
sectors that contain the basis vector $\alpha$, which breaks the $SO(10)$ 
GUT symmetry to the Pati--Salam subgroup. States which arise from sectors 
that do not contain the basis vector $\alpha$ are standard from the point of 
view of the Standard Model charge assignments and grand unification 
representations. Thus, for example, the color triplets appearing in 
eq. (\ref{eqn:vecto}) arise from the vectorial $10$ representation
of the underlying $SO(10)$ GUT symmetry. They are usually termed
leptoquarks in the literature, and are counted as $n_6$ in our analysis.
The experimental constraints 
on these ``standard'' states are not severe and contemporary experiments
are actively seeking their discovery. The experimental constraints on the 
``exotic'' fractionally charged states are far more restrictive. The 
lightest fractionally charged state is necessarily stable and will be
overproduced in a thermal evolution of the early universe. Due to its 
charge it continues to scatter and cannot decouple from the evolving plasma.
Consequently, fractionally charged states must be sufficiently massive
and diluted to avoid constraints from contemporary searches and early universe 
dynamics. It is expected that all non--chiral states
receive mass terms along flat
directions at the high scale, or when the flat directions are lifted
by the SUSY breaking mechanism.

\section{The twisted matter spectrum}\label{projectors}

The counting of spinorials and vectorials is realised by utilising the so
called projectors. Each sector $B^{i}_{pqrs}$, corresponds to a projector
$P^{i}_{pqrs} = 0,1$ which is an entity expressed in terms of GGSO coefficients
and determines the survival or not of a sector. The computational analysis and
manipulation of the projectors becomes more feasible when rewritten in an
analytic form.

\subsection{Observable spinorial states and projectors}

In order to get the particle content for the  representations for the sectors
of \eqref{eqn:obspin} we utilised the following normalisations for the
hypercharge and the electromagnetic charge:
\ba
Y &=& \frac{1}{3} (Q_1 + Q_2 + Q_3) + \frac{1}{2} (Q_4 + Q_5) \\
Q_{em} &=& Y + \frac{1}{2} (Q_4 - Q_5)
 \ea
Where the $Q_{i}$ charges of a state, arise due to $\psi^{i}$ for $i =
1,...,5$.\\
The following table summarises the eigenvalues of the electroweak $SU(2) \times
U(1)$ Cartan generators, in respect to states which fall into the chiral
observable Pati Salam representations:

\begin{tabular}{c|c|c|c|c}
 representation & $\bar{\psi}^{1,2,3}$ & $\bar{\psi}^{4,5}$ & $Y$ & $Q_{em}$ \\
\hline
& $(+,+,+)$ & $(+,+)$ & 1& 1\\
$(\bar{\mathbf{4}},\mathbf{1},\mathbf{2})$ & $(+,+,+)$ & $(-,-)$ & 0& 0\\
& ($+,-,-$)& $(+,+)$ & 1/3& 1/3\\
& ($+,-,-$)& $(-,-)$ & -2/3& -2/3\\
\hline
\hline
& $(-,-,-)$ & $(-,-)$ & -1& -1\\
$(\mathbf{4},\mathbf{1},\mathbf{2})$ & $(-,-,-)$ & $(+,+)$ & 0& 0\\
& ($+,+,-$)& $(-,-)$ & -1/3& -1/3\\
& ($+,+,-$)& $(+,+)$ & 2/3& 2/3\\
\hline
\hline
$(\bar{\mathbf{4}},\mathbf{2},\mathbf{1})$ & $(+,+,+)$ & ($+,-$)& 1/2& 1,0\\
& ($+,-,-$)& ($+,-$)& -1/6& 1/3,-2/3\\
\hline
\hline
$(\mathbf{4},\mathbf{2},\mathbf{1})$ & $(-,-,-)$ & ($+,-$)& -1/2& -1,0\\
& ($+,+,-$)& ($+,-$)& 1/6& -1/3,2/3\\
\end{tabular}\\

In the previous table, $''+''$ and $''-''$  label the contribution of an
oscillator
with fermion number $F = 0$ or $F = -1$  to the degenerate vacuum.
The case of $(+,-,-)$ under $\bar{\psi}^{1,2,3}$ for example, corresponds
to a part of the Ramond vacuum formed by one oscillator with fermion number
$F = 0$ and two oscillators with fermion numbers $F = -1$.
Families and anti-families in the context of these models, can be formed only
if we combine the surviving states of two different sectors:
\ba
\textbf{16} &= (\textbf{4},\textbf{2},\textbf{1}) +
(\bar{\textbf{4}},\textbf{1},\textbf{2}) = N_{4L} + N_{\bar{4}R}\nonumber\\
\bar{\textbf{16}}& =  (\textbf{4},\textbf{1},\textbf{2}) +
(\bar{\textbf{4}},\textbf{2},\textbf{1}) = N_{4R} + N_{\bar{4}L}
\ea
A phenomenologically viable model, must of course consist of only 3 families:
\ba
 N_{4L}  -  N_{\bar{4}L} = N_{\bar{4}R} -  N_{4R}  = 3
\ea
In order to be able to distinguish between $N_{4L},N_{\bar{4}L}, N_{\bar{4}R}$
and $N_{4R}$, one has to define Representation Operators that will determine
the representations in which the states of each observable sector, will fall
into.The operators $X_{pqrs}^{{i}_{SU(4)}} = \pm 1$ that define the $SU(4)$
chirality $(\mathbf{4}$ or  $\mathbf\bar{{4}})$ for $B^{1}_{pqrs}$ ,
$B^{2}_{pqrs}$ and $B^{3}_{pqrs}$ respectively are:
\begin{eqnarray}
X_{pqrs}^{(1)_{SU(4)}} & = &
C\binom{B^{(1)}_{pqrs}}{S + b_{2} + \alpha + (1-r)e_{5} + (1-s)e_{6}}\nonumber\\
X_{pqrs}^{(2)_{SU(4)}} & = &
C\binom{B^{(2)}_{pqrs}}{S + b_{1} + \alpha + (1-r)e_{5} + (1-s)e_{6}}\\
X_{pqrs}^{(3)_{SU(4)}} & = &
C\binom{B^{(3)}_{pqrs}}{S + b_{2} + \alpha + (1-p)e_{1} + (1-q)e_{2}}\nonumber
\end{eqnarray}
The  representation operators $X_{pqrs}^{{(i)}_{SU(2)_{L/R}}} = \pm1$ determine
the $SU(2)_{L/R}$ representations $((\textbf{1,2})$ or  $(\textbf{2,1}))$ for
$B^{(1)}_{pqrs}$ , $B^{(2)}_{pqrs}$ and $B^{(3)}_{pqrs}$ respectively.In the
following expressions $V_i = S + b_{i} + \alpha + x$.

\begin{eqnarray}
X_{pqrs}^{(1)_{SU(2)_{L/R}}} & = &
C\binom{B^{(1)}_{pqrs}}{V_1  + (1-p)e_{3} + (1-q)e_{4} + (1-r)e_{5} +
(1-s)e_{6}}\nonumber\\
X_{pqrs}^{(2)_{SU(2)_{L/R}}} & = &
C\binom{B^{(2)}_{pqrs}}{V_2  + (1-p)e_{1} + (1-q)e_{2} + (1-r)e_{5} +
(1-s)e_{6}}\\
X_{pqrs}^{(3)_{SU(2)_{L/R}}} & = &
C\binom{B^{(3)}_{pqrs}}{V_3 + (1-p)e_{1} + (1-q)e_{2} + (1-r)e_{3} +
(1-s)e_{4}}\nonumber
\end{eqnarray}

The explicit expressions for the 48 projectors related to the observable chiral
matter are:
\begin{eqnarray}
P_{pqrs}^{(1)}&=&
\frac{1}{4}\,\left(1-c\binom{e_1}{B_{pqrs}^{(1)}} \right)\,
\cdot\left(1-c\binom{e_2}{B_{pqrs}^{(1)}}\right)\,\nonumber\\
&&\cdot\frac{1}{4}\left(1-c\binom{z_1}{B_{pqrs}^{(1)}}\right)\,
\cdot\left(1-c\binom{z_2}{B_{pqrs}^{(1)}}\right) \nonumber\\
P_{pqrs}^{(2)}&=&
\frac{1}{4}\,\left(1-c\binom{e_3}{B_{pqrs}^{(2)}}\right)\,
\cdot\left(1-c\binom{e_4}{B_{pqrs}^{(2)}}\right)\,\nonumber\\
&&\cdot\frac{1}{4}\left(1-c\binom{z_1}{B_{pqrs}^{(2)}}\right)\,
\cdot\left(1-c\binom{z_2}{B_{pqrs}^{(2)}}\right)\\
P_{pqrs}^{(3)}&=&
\frac{1}{4}\,\left(1-c\binom{e_5}{B_{pqrs}^{(3)}}\right)\,
\cdot\left(1-c\binom{e_6}{B_{pqrs}^{(3)}}\right)\, \nonumber\\
&&\cdot\frac{1}{4}\left(1-c\binom{z_1}{B_{pqrs}^{(3)}}\right)\,
\cdot\left(1-c\binom{z_2}{B_{pqrs}^{(3)}}\right) \nonumber
\end{eqnarray}
Using the appropriate formalism these projectors can be expressed as a system
of linear equations with $p$, $q$, $r$ and $s$ as unknowns. The solutions of a
specific system of equations, yield the different combinations of $p$, $q$,
$r$, $s$ for which sectors survive the GSO projections.This formalism is more
suitable and much more flexible for a computer-oriented analysis. In order to
achieve the transition to this formalism, the following notation is
introduced
\ba
\cc{a_i}{a_j}=e^{i \pi \oo{a_i}{a_j}}\,\ ,\  \oo{a_i}{a_j}=0,1, 
\ea
where $a_i$ and $a_j$ refer to the basis vectors, and the GGSO projection 
coefficients are defined in eq. \ref{eq:gso}.
The new expression implies properties which can be easily derived after
performing standard algebraic methods involving the GGSO coefficients
\ba
\oo{a_i}{a_j+a_k}&=&\oo{a_i}{a_j}+\oo{a_i}{a_k} \ ,
\forall\ a_{i}: \{\psi^\mu\}\cap a_i=\O
\\
\oo{a_i}{a_j}&=&\oo{a_j}{a_i} \ ,
\forall\ a_{i},a_{j}: a_i\cdot a_j=0\ {\rm mod}\ 4
\ea
where $\# (a_i\cdot a_j)\equiv\# \left[a_i\cup a_j - a_i\cap a_j\right]$.\\

The analytic expressions for each different projector $P^{1,2,3}_{pqrs}$
respectively, are given in a matrix form $\Delta^{i}W^{i} = Y^{i}$.
\ba
\left(
\begin{array}{cccc}
\oo{e_1}{e_3}&\oo{e_1}{e_4}&\oo{e_1}{e_5}&\oo{e_1}{e_6}\\
\oo{e_2}{e_3}&\oo{e_2}{e_4}&\oo{e_2}{e_5}&\oo{e_2}{e_6}\\
\oo{z_1}{e_3}&\oo{z_1}{e_4}&\oo{z_1}{e_5}&\oo{z_1}{e_6}\\
\oo{z_2}{e_3}&\oo{z_2}{e_4}&\oo{z_2}{e_5}&\oo{z_2}{e_6}
\end{array}
\right)
\left(
\begin{array}{c}
p\\
q\\
r\\
s\\
\end{array}
\right)
=
\left(
\begin{array}{c}
\oo{e_1}{b_1}\\
\oo{e_2}{b_1}\\
\oo{z_1}{b_1}\\
\oo{z_2}{b_1}
\end{array}
\right) \nonumber
\ea
\vspace{-0.2cm}
\ba
\left(
\begin{array}{cccc}
\oo{e_3}{e_1}&\oo{e_3}{e_2}&\oo{e_3}{e_5}&\oo{e_3}{e_6}\\
\oo{e_4}{e_1}&\oo{e_4}{e_2}&\oo{e_4}{e_5}&\oo{e_4}{e_6}\\
\oo{z_1}{e_1}&\oo{z_1}{e_2}&\oo{z_1}{e_5}&\oo{z_1}{e_6}\\
\oo{z_2}{e_1}&\oo{z_2}{e_2}&\oo{z_2}{e_5}&\oo{z_2}{e_6}
\end{array}
\right)
\left(
\begin{array}{c}
p\\
q\\
r\\
s\\
\end{array}
\right)
=
\left(
\begin{array}{c}
\oo{e_3}{b_2}\\
\oo{e_4}{b_2}\\
\oo{z_1}{b_2}\\
\oo{z_2}{b_2}
\end{array}
\right)
\ea
\vspace{-0.2cm}
\ba
\left(
\begin{array}{cccc}
\oo{e_5}{e_1}&\oo{e_5}{e_2}&\oo{e_5}{e_3}&\oo{e_5}{e_4}\\
\oo{e_6}{e_1}&\oo{e_6}{e_2}&\oo{e_6}{e_3}&\oo{e_6}{e_4}\\
\oo{z_1}{e_1}&\oo{z_1}{e_2}&\oo{z_1}{e_3}&\oo{z_1}{e_4}\\
\oo{z_2}{e_1}&\oo{z_2}{e_2}&\oo{z_2}{e_3}&\oo{z_2}{e_4}
\end{array}
\right)
\left(
\begin{array}{c}
p\\
q\\
r\\
s\\
\end{array}
\right)
=
\left(
\begin{array}{c}
\oo{e_5}{b_3}\\
\oo{e_6}{b_3}\\
\oo{z_1}{b_3}\\
\oo{z_2}{b_3}
\end{array}
\right) \nonumber
\ea

The corresponding algebraic expressions for the states from the remaining sectors
above are given in appendix \ref{appendixa}.
We note that although the hidden sector 
states can play a crucial phenomenological role, like for
example in SUSY breaking, their classification is not done
in the analysis here, which focuses exclusively on states that are charged under the 
Standard Model group factors. Our aim in the present paper in particular
is the classification in respect to the fractionally charged states.
Experimental observations demand that the low energy exotic states should be
truncated from the spectrum or accommodate heavy mass. The projectors
shown in appendix \ref{appendixa} are
crucial in this regard since their values
determines the number of surviving exotic
representations in each model.

\section{The four dimensional gauge group}

The untwisted spectrum is common in all the Pati--Salam vacua that
we classify. The models differ by the states that arise from the sectors
in eq. (\ref{ggsectors})
In our classification method the GGSO projections
are encoded in algebraic equations that depend on the GGSO projection
coefficients, and are applied to all the sectors listed in section
\ref{analysis}.

If the gauge bosons of a sector transform under a subgroup of the
Neveu--Schwarz gauge group, the NS gauge group is enhanced.
We restrict the class of vacua to the cases without enhancement.
We therefore find the conditions under which the gauge bosons of
a specific sector survive. Below we present the type of
enhancements that can occur from different sectors,
assuming that only one set of conditions is satisfied in each distinct case.

\subsection{Enhancements of the Observable gauge group}

\begin{itemize}
\item
$x = \{\bar{\eta}^{123}, \bar{\psi}^{12345}\}$ is the only sector
which can enlarge the observable gauge group.
Enhancement takes place when the following conditions are satisfied\\

\begin{tabular}{| l | l |}
\hline
Enhancement  conditions&Resulting Enhancement\\
\hline
$(x | e_{i}) = (x | z_n) = 0 $&$ SU(4)_{obs}\times SU(2)_{L/R}
\times U(1)' \rightarrow SU(6)$\\
\hline
\end{tabular}\\

The pre-stated conditions hold
for all $i = 1, ... , 6 , \, n = 1, 2$, and $U(1)'$
is a linear combination of the $U(1)_{i}$ where $i = 1,2,3$.
In the case that any of the previous conditions is not satisfied,
the enlargement of the gauge group is not possible.
\end{itemize}

\subsection{Enhancements of the Hidden gauge group}

\begin{itemize}

\item $z_1 + z_2 = \{ \bar{\phi}^{12345678}\}$ is
the only sector that enlarges only
the hidden gauge group when all of the following conditions are met:\\

\begin{tabular}{| l | l |}
\hline
Enhancement  conditions&Resulting Enhancement\\
\hline
$(e_i | z_1 + z_2) = (b_k | z_1 + z_2) = 0 $&$
SU(2)_{1/2}\times SU(2)_{3/4}  \times SO(8)_{hid} \rightarrow SO(12) $\\
$ \forall \, i = 1, ... , 6 , \, k = 1, 2$& \\
\hline
\end{tabular}

\end{itemize}

\subsection{Mixed gauge group enhancements}

Parts of the observable and hidden gauge group can be enhanced
simultaneously in the following cases.


\begin{itemize}

\item $\alpha + z_{1} + z_{2} = \{\bar{\psi}^{45},
\bar{\phi}^{34},\bar{\phi}^{5678}\}$ \\

\begin{tabular}{| l | l |}
\hline
Enhancement Conditions&Resulting Enhancement\\

\hline
$(e_{i} | \alpha + z_{1} + z_{2}) =  0
$&$SU(2)_{L/R}\times SU(2)_{3/4}\times SO(8)_{hid}$\\
$(b_1 | \alpha + z_{1} +  z_2) = (b_2 | \alpha +
z_{1} +  z_2) = (\alpha | \alpha + z_{1} + z_{2})
$&$ \rightarrow SO(12)$\\
$(1 | \alpha + z_1 +z_{2})~ = 1 + (b_k |
\alpha + z_{1} +  z_{2}) $&\\
\hline
\end{tabular}\\

The conditions of the previous table hold  for all  $i = 1, ... , 6 $\\

\item $\alpha + x + z_1 = \{\bar{\eta}^{123},
\bar{\psi}^{123},\bar{\phi}^{34}\}$\\

\begin{tabular}{| l | l |}
\hline
Enhancement Conditions&Resulting Enhancement\\

\hline
$(e_{i} | \alpha + x + z_1) = (z_2 | \alpha + x + z_1) =  0
$&$SU(4)_{obs}\times SU(2)_{1/2} \times U(1)^{'}\rightarrow SU(6)$\\
\hline
\end{tabular}\\

The conditions above hold for all $i = 1, ... , 6 $\\

\item $\alpha + x = \{\bar{\eta}^{123}, \bar{\psi}^{123},\bar{\phi}^{12}\}$ \\

\begin{tabular}{| l | l |}
\hline
Enhancement Conditions&Resulting Enhancement\\

\hline
$(e_{i} | \alpha + x) = (z_2 | \alpha + x) =
0 \, , \quad \forall \, i = 1, ... , 6
$&$SU(4)_{obs}\times SU(2)_{1/2} \times U(1)^{'}
\rightarrow SU(6)$\\
$(z_1 | \alpha + x) = (\alpha | \alpha + x) $&\\
\hline
\end{tabular}\\

\item $\alpha + z_{2} = \{\bar{\psi}^{45}, \bar{\phi}^{12},\bar{\phi}^{5678}\}$ \\

\begin{tabular}{| l | l |}
\hline
Enhancement Conditions &Resulting Enhancement\\

\hline
$(e_{i} | \alpha + z_{2}) =  0 \, , \quad
\forall \, i = 1, ... , 6
$&$SU(2)_{L/R}\times SU(2)_{1/2} \times SO(8)_{hid}
\rightarrow SO(12)$\\
$(b_1 | \alpha + z_2) = (b_2 | \alpha + z_2)$ &\\
$(b_k | \alpha + z_2) + (z_1 | \alpha + z_2) =
(\alpha | \alpha + z_{2})$& \\

\hline
\end{tabular}\\

\item $z_1 = \{ \bar{\phi}^{1234}\}$ produces the following enhancements:\\

\begin{tabular}{| l | l |}
\hline
Survival Conditions & Resulting Enhancement\\

\hline
$(e_i | z_1) = (z_2 | z_1) = 0$&$
SU(4)_ {obs}\times SU(2)_{1/2}\times SU(2)_{3/4}\rightarrow SO(10)$\\
$(b_k | z_1) = 1$& \\
\hline
$(e_i | z_1) = (z_2 | z_1) = 0$&$
SU(2)_ {L}\times SU(2)_{R}\times SU(2)_{2/1}\times
SU(2)_{4/3}\rightarrow SO(8)$\\
$(b_k | z_1) = 1$&\\
\hline
$(e_i | z_1) = (z_2 | z_1) =(b_2 | z_1) =
0$&$ SU(2)_{1/2} \times SU(2)_{3/4} \times U(1) \rightarrow SO(6)$\\
$(b_1 | z_1) = 1$&\\
\hline
$(e_i | z_1) = (z_2 | z_1) =(b_1 | z_1) =
0$&$ SU(2)_{1/2} \times SU(2)_{3/4}\times U(1)\rightarrow SO(6)$\\
$(b_2 | z_1) = 1$&  \\
\hline
$(e_i | z_1) = (z_2 | z_1) =(b_k | z_1) =
0$&$ SU(2)_{1/2} \times SU(2)_{3/4} \times U(1) \rightarrow SO(6)$\\
\hline
$(e_j | z_1) = (z_2 | z_1) = 0 $&$
SU(2)_{1/2} \times SU(2)_{3/4}  \rightarrow SO(5)$\\
$(e_i | z_1) = 1$&\\
AND&\\
$(b_1 | z_1) = 0, (b_2 | z_1) = 1, i = 1,2$&\\
or&\\
$(b_1 | z_1) = 1, (b_2 | z_1) = 0, i = 3,4$&\\
or&\\
$(b_1 | z_1) = 1, (b_2 | z_1) = 1, i = 5,6$&\\
\hline
$(e_j | z_1) = (z_2 | z_1) = 0 $&$ SU(2)_{1/2}
\times SU(2)_{3/4}  \rightarrow SO(5)$\\
$(e_i | z_1) = 1$&\\
$(b_k | z_1) = 0$&\\
\hline
$(e_{i} | z_1) = (b_k | z_1) = 0$&$SU(2)_{1/2}
\times SU(2)_{3/4} \times SO(8)_{hid}\rightarrow SO(12)$\\
$(z_2 | z_1) = 1$&\\
\hline
\end{tabular}\\

The relations above, hold for all $i,j = 1,....6$
where $i \ne j$ and $k = 1,2$. We note that while $z_2$
produces two cases in which only the hidden $SO(8)$ gauge group is 
enhanced to $SO(9)$, in other cases it leads to enhancements that mix
the hidden and observable gauge groups. 


\item $z_2 = \{ \bar{\phi}^{5678}\}$ can generate
enhancements in the following cases:\\

\begin{tabular}{| l | l |}
\hline
Survival  Conditions  & Resulting Enhancement\\

\hline
$(e_i | z_2) = (z_1 | z_2) = (\alpha | z_2) = 0$&$
SU(4)_ {obs}\times SO(8)_{hid} \rightarrow SO(14)$\\
$(b_k | z_2) = 1$&\\
\hline
$(e_i | z_2) = (z_1 | z_2) = 0$&$
SU(2)_ {L}\times SU(2)_{R} \times SO(8)_{hid}\rightarrow SO(12)$\\
$(b_k | z_2) =(\alpha | z_2) = 1$&\\
\hline
$(e_i | z_2) = (z_1 | z_2) =(b_2 | z_2) = (\alpha | z_2) =
0$&$U(1)\times SO(8)_{hid} \rightarrow SO(10)$\\
$(b_1 | z_2) = 1$&\\
\hline
$(e_i | z_2) = (z_1 | z_2) =(b_1 | z_2) =(\alpha | z_2) =
0$&$ U(1)\times SO(8)_{hid} \rightarrow SO(10)$\\
$(b_2 | z_2) = 1$&\\
\hline
$(e_i | z_2) = (z_1 | z_2) =(b_k | z_2) = (\alpha | z_2) =
0$&$ U(1)\times SO(8)_{hid} \rightarrow SO(10)$\\
\hline
$(e_j | z_2) = (z_1 | z_2) = (\alpha | z_2) = 0 $&$
SO(8)_{hid} \rightarrow SO(9)$\\
$(e_i | z_2) = 1$&\\
AND&\\
$(b_1 | z_2) = 0, (b_2 | z_2) = 1, i = 1,2$&\\
or&\\
$(b_1 | z_2) = 1, (b_2 | z_2) = 0, i = 3,4$&\\
or&\\
$(b_1 | z_2) = 1, (b_2 | z_2) = 1, i = 5,6$&\\
\hline
$(e_j | z_2) = (z_1 | z_2) =(b_k | z_2) = (\alpha | z_2) = 0
$&$ SO(8)_{hid} \rightarrow SO(9)$\\
$(e_i | z_2) = 1$&\\
\hline
$(e_i | z_2) = (b_k | z_2) = 0$&$SO(4)_{1} \times SO(8)_{hid}
 \rightarrow SO(12)$\\
$(\alpha | z_2) = (z_1 | z_2) = 1$&\\
\hline
$(e_i | z_2) = (b_k | z_2) = (\alpha | z_2) =
0$&$SO(4)_{2} \times SO(8)_{hid} \rightarrow SO(12)$\\
$(z_1 | z_2) = 1$&\\
\hline
\end{tabular}\\

The relations above, hold for all $i,j = 1,....6$ where
$i \ne j$ and $k = 1,2$

\newpage

\item $\alpha = \{\bar{\psi}^{45} \bar{\phi}^{12}\}$
can also present numerous potential enhancements.\\

\begin{tabular}{| l | l |}
\hline
 Survival Conditions & Resulting  Enhancement\\

\hline
$(e_i | \alpha) = (z_2 | \alpha) = 0$&$
SU(4)_ {obs}\times SU(2)_{L/R}\times SU(2)_{1/2}\rightarrow SO(10)$\\
$(b_1 | \alpha) = (b_2 | \alpha) $&AND\\
$(1 | \alpha) = 1 + (b_k | \alpha) +
(z_1 | \alpha) $&$SU(2)_{L/R}\times SU(2)_{1/2}
\times SU(2)_3 \times SU(2)_4\  \rightarrow SO(8)$\\\\
\hline
$(e_i | \alpha) = (z_2 | \alpha) = 0
$&$U(1)\times SU(2)_{L/R}\times SU(2)_{1/2} \rightarrow SO(6)$\\
$(b_1 | \alpha) = 1 + (b_2 | \alpha) $&\\
$(1 | \alpha) = (b_1 | \alpha) + (z_1 | \alpha) $&\\
\hline
$(e_i | \alpha) = (z_2 | \alpha) = 0
$&$U(1)\times SU(2)_{L/R} \times SU(2)_{1/2}\rightarrow SO(6)$\\
$(b_2 | \alpha) = 1 + (b_1 | \alpha) $& \\
$(1 | \alpha) = (b_2 | \alpha) + (z_1 | \alpha) $&\\
\hline
$(e_i | \alpha) = (z_2 | \alpha) = 0
$&$U(1)\times SU(2)_{L/R}\times SU(2)_{1/2} \rightarrow SO(6)$\\
$(b_1| \alpha) = (b_2 | \alpha) $&\\
$(1 | \alpha) = (b_2 | \alpha) + (z_1 | \alpha) $&\\
\hline
$(e_j | \alpha) = (z_2 | \alpha) = 0
$&$ SU(2)_{L/R}\times SU(2)_{1/2}  \rightarrow SO(5)$\\
$(e_i | \alpha) = 1$&\\
AND&\\
$(b_1 | \alpha) = 1+ (b_2 | \alpha)$ and&\\
$(1 | \alpha) = (b_1 |\alpha) + (z_1 | \alpha),~~~~~i = 1,2$&\\
or&\\
$(b_1 | \alpha) = 1+ (b_2 | \alpha)$ and&\\
$(1 | \alpha) = (b_2 |\alpha) + (z_1 | \alpha),~~~~~i = 3,4$&\\
or&\\
$(b_1 | \alpha) =  (b_2 | \alpha)$ and&\\
$(1 | \alpha) = 1 + (b_k |\alpha) + (z_1 | \alpha),i = 5,6$&\\
\hline
$(e_j | \alpha) = (z_2 | \alpha) = 0
$&$ SU(2)_{L/R}\times SU(2)_{1/2}  \rightarrow SO(5)$\\
$(e_i | \alpha) = 1$&\\
$(b_1 | \alpha) = (b_2 | \alpha)$&\\
$(1 | \alpha) = (b_k | \alpha) + (z_1 | \alpha)$&\\
\hline
$(e_i | \alpha) =  0$&$SU(2)_{L/R} \times
SU(2)_{1/2}  \times SO(8)_{hid} \rightarrow SO(12)$\\
$(z_2 | \alpha) = 1$&\\
$(b_1 | \alpha) =  (b_2 | \alpha)$&\\
$(1 | \alpha) =  (b_k | \alpha) + (z_1 | \alpha)$&\\
\hline
\end{tabular}\\

The relations above, hold for all $i,j = 1,....6$ where $i \ne j$ and $k = 1,2$

\newpage

\item $\alpha  + z_{1} = \{\bar{\psi}^{45} \bar{\phi}^{34}\}$ gives rise to
enhancements in the following occasions:\\

\begin{tabular}{|l|l|}
\hline
Survival Conditions  & Resulting Enhancement\\

 \hline
$(e_i | \alpha + z_1) = (z_2 | \alpha + z_1) =
0$&$ SU(4)_ {obs}\times SU(2)_{L/R} \times SU(2)_{3/4}$\\
$(b_1 | \alpha + z_1) = (b_2 | \alpha + z_1) $&$\rightarrow SO(10)$\\
$(\alpha | \alpha + z_1) = 1 + (b_k | \alpha + z_1) $&AND\\

&$SU(2)_{L/R}\times SO(4)_{1}\times SU(2)_{3/4}$\\
&$\rightarrow SO(8)$\\
\hline
$(e_i | \alpha + z_1) = (z_2 | \alpha + z_1) = 0
$&$U(1)\times SU(2)_{L/R} \times SU(2)_{3/4}$\\
$1+ (b_1 | \alpha + z_1) =  (b_2 | \alpha + z_1)
= (\alpha | \alpha + z_1) $&$ \rightarrow SO(6)$\\
\hline
$(e_i | \alpha+ z_1) = (z_2 | \alpha+ z_1)  =
0$&$U(1)\times SU(2)_{L/R}\times SU(2)_{3/4}$\\
$1 + (b_2 | \alpha + z_1) =  (b_1 | \alpha + z_1) =
(\alpha | \alpha + z_1) $&$ \rightarrow SO(6)$\\
\hline
$(e_i | \alpha + z_1) = (z_2 | \alpha + z_1) = 0
$&$U(1)\times SU(2)_{L/R}\times SU(2)_{3/4}$\\
$(b_1| \alpha + z_1) = (b_2 | \alpha) =
(\alpha | \alpha + z_1)$ &$\rightarrow SO(6)$\\
\hline
$(e_j | \alpha + z_1) = (z_2 | \alpha + z_1) = 0
$&$ SU(2)_{L/R}\times SU(2)_{3/4}  \rightarrow SO(5)$\\
$(e_i | \alpha + z_1) = 1$&\\
AND&\\
$(b_1 | \alpha + z_1) = 1+ (b_2 | \alpha + z_1) =
(\alpha | \alpha + z_1),i = 1,2$&\\
or&\\
$(b_1 | \alpha + z_1) = 1+ (b_2 | \alpha + z_1) =
1 + (\alpha| \alpha + z_1),i = 3,4$&\\
or&\\
$(b_1 | \alpha + z_1) =  (b_2 | \alpha + z_1) =
1+  (\alpha | \alpha + z_1),i = 5,6$&\\
\hline
$(e_j | \alpha + z_1) = (z_2 | \alpha + z_1) =
0 $&$ SU(2)_{L/R}\times SU(2)_{1/2} \rightarrow SO(5)$\\
$(e_i | \alpha + z_1) = 1$&\\
$(b_1 | \alpha + z_1) = (b_2 | \alpha + z_1)$&\\
$(1 | \alpha + z_1) = (b_k | \alpha + z_1) + (z_1 | \alpha + z_1)$&\\

\hline
$(e_i | \alpha + z_1) =  0$&$SU(2)_{L/R} \times SU(2)_{3/4}\times SO(8)$\\
$(z_2 | \alpha + z_1) = 1$&$ \rightarrow SO(12)$\\
$(b_1 | \alpha + z_1) = (b_2 | \alpha + z_1) = (\alpha | \alpha + z_1) $&\\
\hline
\end{tabular}\\


\end{itemize}

\section{Results}

Using the algebraic expressions presented in the previous
sections we can analyse
the entire massless spectrum for a given choice of
GGSO projection coefficients
that completely specify a specific string model.
These formulas are inputted into
a computer program which is used to scan the space
of string vacua produced by
random generation of the one--loop GGSO projection
coefficients. The number of
possible configurations is $2^{51}\sim 10^{15}$,
which is too large for a complete
classification. For this reason a random generation
algorithm is
utilised\footnote{We note that analysis of large sets
of string vacua has also
been performed by other groups \cite{statistical}.},
and the characteristics of the model for each set of
random GGSO projection
coefficients are extracted. In this manner a model
with some desired phenomenological
criteria can be fished from the sample generated.
In ref. \cite{exophobic}
this procedure was followed and produced a three generation Pati--Salam
string model that does not contain any exotic massless states with fractional
electric charge. In this paper we use this methodology to classify the
Pati--Salam free fermionic string models with respect to some phenomenological
criteria. The observable sector of a heterotic--string Pati--Salam model is
characterized by 9 integers
$\left(n_g,k_L,k_R,n_6,n_h,n_4,n_{\bar{4}},n_{2L},n_{2R}\right)$, where
\begin{align}
&n_{4L}-n_{\bar{4}L}=n_{\bar{4}R}-n_{4R}=n_g=\text{\# of generations}\nn\\
&n_{\bar{4}L}=k_L=\text{\# of non chiral left pairs}\nn\\
&n_{4R}=k_R=\text{\# of non chiral right pairs}\nn\\
&n_6=\text{\# of } ({\textbf 6},{\textbf 1},{\textbf 1})\nn\\
&n_h=\text{\# of } ({\textbf 1},{\textbf 2},{\textbf 2})\nn\\
&n_4=\text{\# of } ({\textbf 4},{\textbf 1},{\textbf 1})\text{ (exotic)}\nn\\
&n_{\bar{4}}=\text{\# of } ({\bar{\textbf 4}},{\textbf 1},{\textbf 1})\text{ (exotic)}\nn\\
&n_{2L}=\text{\# of } ({\textbf 1},{\textbf 2},{\textbf 1})\ \text{ (exotic)}\nn\\
&n_{2R}=\text{\# of } ({\textbf 1},{\textbf 1},{\textbf 2}) \text{ (exotic)}\nn
\end{align}
Using the methodology outlined in section
\ref{projectors} we obtain analytic formulas for all these quantities.
The spectrum of a viable Pati--Salam heterotic string model should
have $n_g=3$,
\begin{align}
&n_g=3~~~\text{three light chiral of generations}\nn\\
&k_L\ge0~~~\text{heavy mass can be generated for non chiral pairs}\nn\\
&k_R\ge1~~~\text{at least one Higgs pair to break the PS symmetry}\nn\\
&n_6\ge1~~~\text{at least one required for missing partner mechanism }\nn\\
&n_h\ge1~~~\text{at least one light Higgs bi--doublet}\nn\\
&n_4=n_{\bar{4}}\ge0~~~\text{heavy mass can be generated for vector--like
exotics}\nn\\
&n_{2L}=0\text{mod}2~~~\text{heavy mass can be generated for vector--like
exotics}\nn\\
&n_{2R}=0\text{mod}2~~~\text{heavy mass can be generated for vector--like
exotics}\nn
\end{align}

A minimal model which is free of exotics has
$k_L=0$, $k_R=1$,
$n_6=1$, $n_h=3$, $n_4=n_{\bar{4}}=0$, $n_{2L}=0$ and $n_{2R}=0$.
The model given by the following GGSO coefficients matrix :

\beq [v_i|v_j] = e^{i\pi (v_i|v_j)} \eeq

\beq \label{BigMatrix}  (v_i|v_j)\ \ =\ \ \bordermatrix{
        & 1& S&e_1&e_2&e_3&e_4&e_5&e_6&b_1&b_2&z_1&z_2&\alpha\cr
 1  & 1& 1& 1& 1& 1& 1& 1& 1& 1& 1& 1& 1 &0\cr
S  & 1& 1& 1& 1& 1& 1& 1& 1& 1& 1& 1& 1& 1\cr
e_1& 1& 1& 0& 1& 0& 0& 0& 0& 0& 0& 0& 1& 1\cr
e_2& 1& 1& 1& 0& 1& 1& 1& 0& 1& 0& 1& 1& 0\cr
e_3& 1& 1& 0& 1& 0& 1& 1& 0& 0& 1& 0& 0& 0\cr
e_4& 1& 1& 0& 1& 1& 0& 1& 0& 0& 1& 0& 0& 1\cr
e_5& 1& 1& 0& 1& 1& 1& 0& 1& 1& 0& 0& 0& 0\cr
e_6& 1& 1& 0& 0& 0& 0& 1& 0& 0& 0& 0& 1& 1\cr
b_1& 1& 0& 0& 1& 0& 0& 1& 0& 1& 1& 0& 0& 1\cr
b_2& 1& 0& 0& 0& 1& 1& 0& 0& 1& 1& 1& 1& 0\cr
z_1& 1& 1& 0& 1& 0& 0& 0& 0& 0& 1& 1& 1& 1\cr
z_2& 1& 1& 1& 1& 0& 0& 0& 1& 0& 1& 1& 1& 0\cr
\alpha& 0& 1& 1& 0& 0& 1& 0& 1& 0& 1& 0& 0& 0\cr
  }
\eeq

\noindent
was presented in ref. \cite{exophobic} and produces the desired spectrum.
The twisted sectors in this model produce
three chiral generations; one pair of heavy Higgs states;
one light Higgs bi--doublet; one vector sextet of $SO(6)$; and is completely
free
of massless exotic fractionally charged states. Additionally the model contains
three pairs of untwisted $SO(6)$ sextets, which can obtain string
scale mass
along flat directions. The full massless spectrum of this model was presented in ref.
\cite{exophobic}.

\begin{figure}[!ht]
\centering
\includegraphics[width=\textwidth]{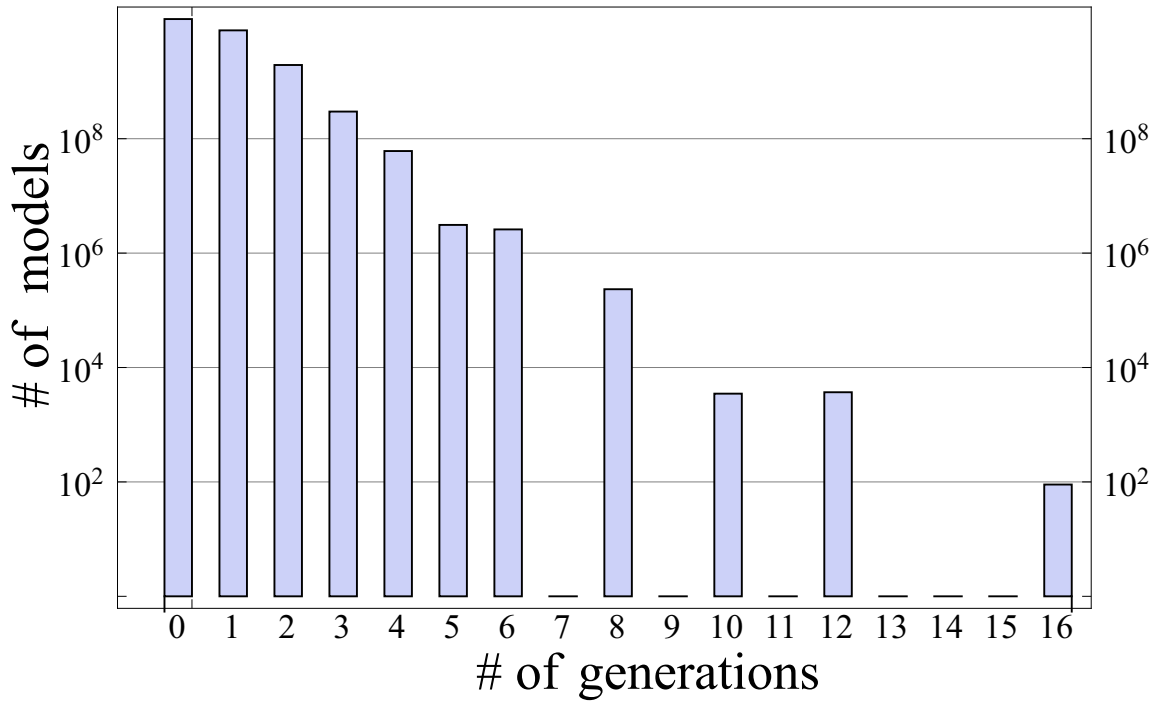}
\caption{\label{gen_distribution}
\it Number of models versus number of generations
($n_g$) in a random sample of $10^{11}$ GGSO configurations.}
\end{figure}

\begin{figure}[!ht]
\centering
\includegraphics[width=\textwidth]{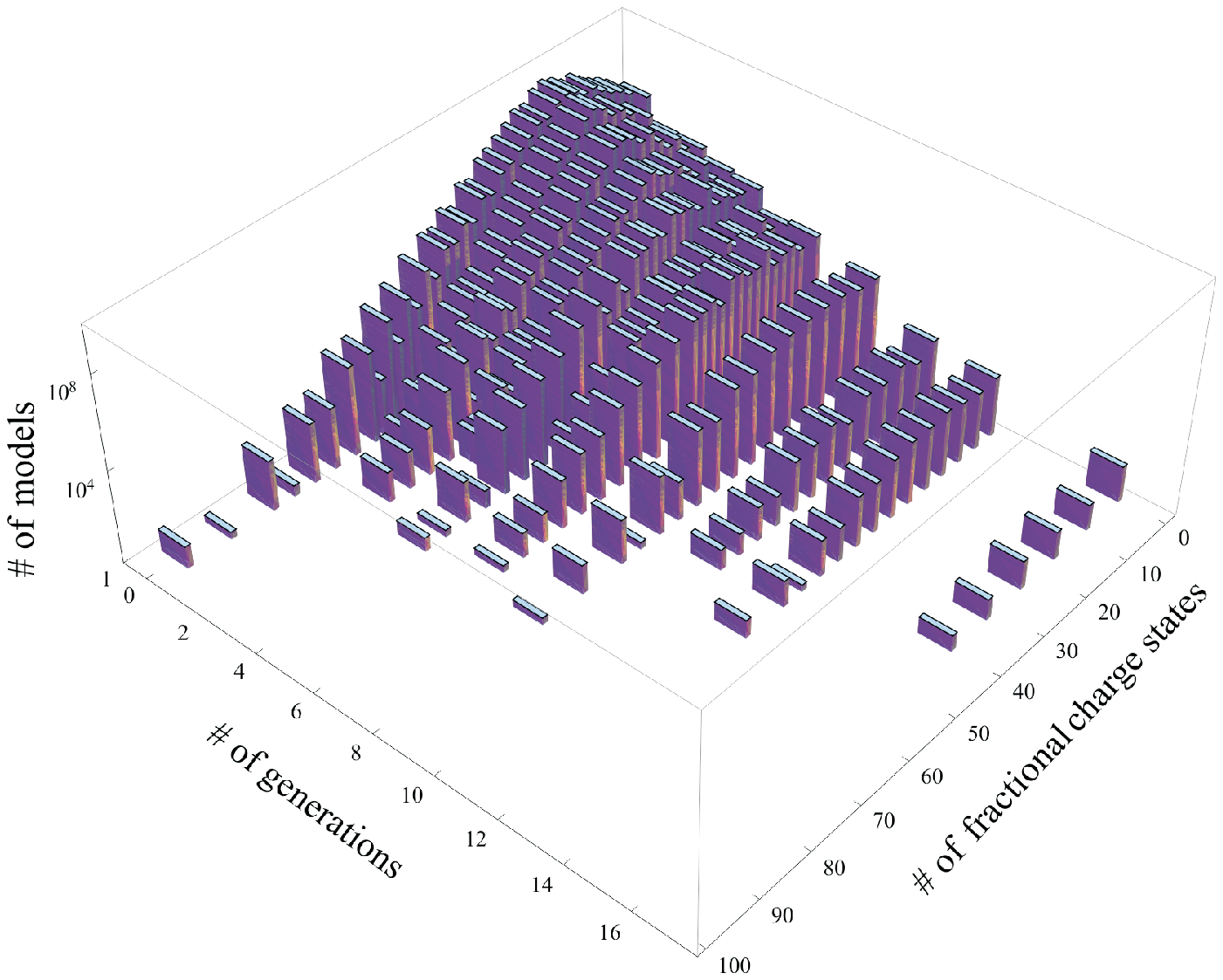}
\caption{\label{gen_frac}
\it Number of  models versus number of generations
($n_g$)  and total number of exotic multiplets in a random sample of
$10^{11}$ GGSO configurations.}
\end{figure}

\begin{figure}[!ht]
\centering
\includegraphics[width=\textwidth]{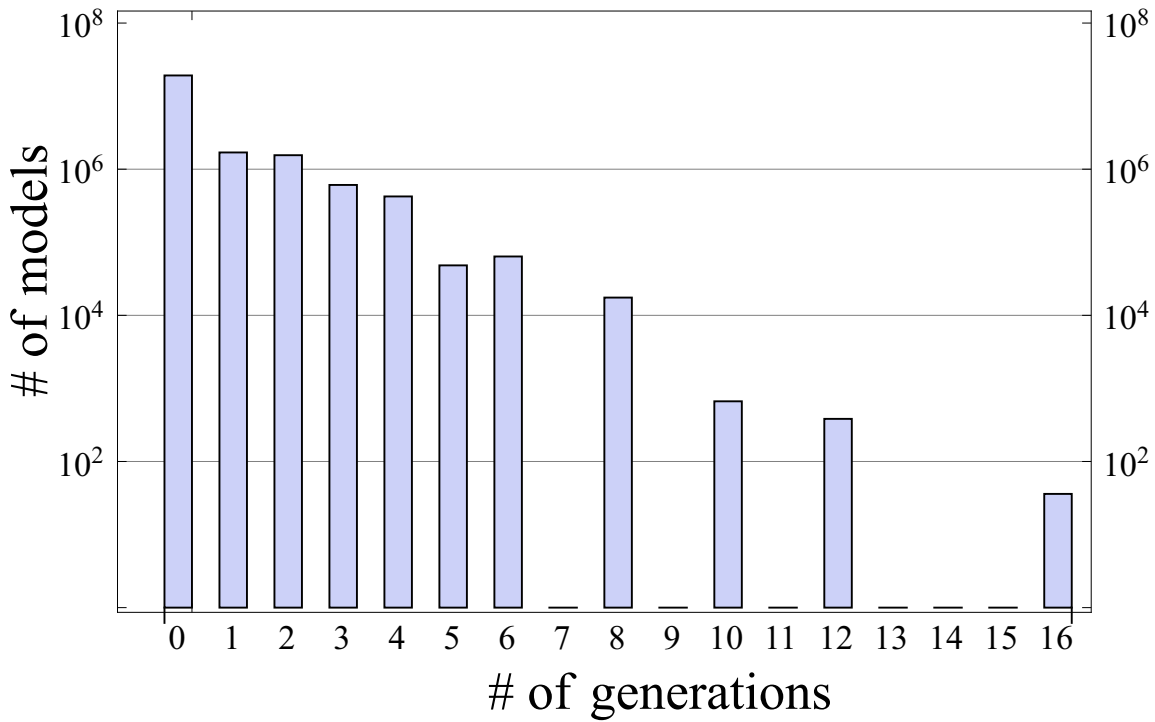}
\caption{\label{gen_no_fracs}
\it Number of exotic free models versus number of generations
($n_g$) in a random sample of $10^{11}$ GGSO configurations.}
\end{figure}

\begin{figure}[!ht]
\centering
\includegraphics[width=\textwidth]{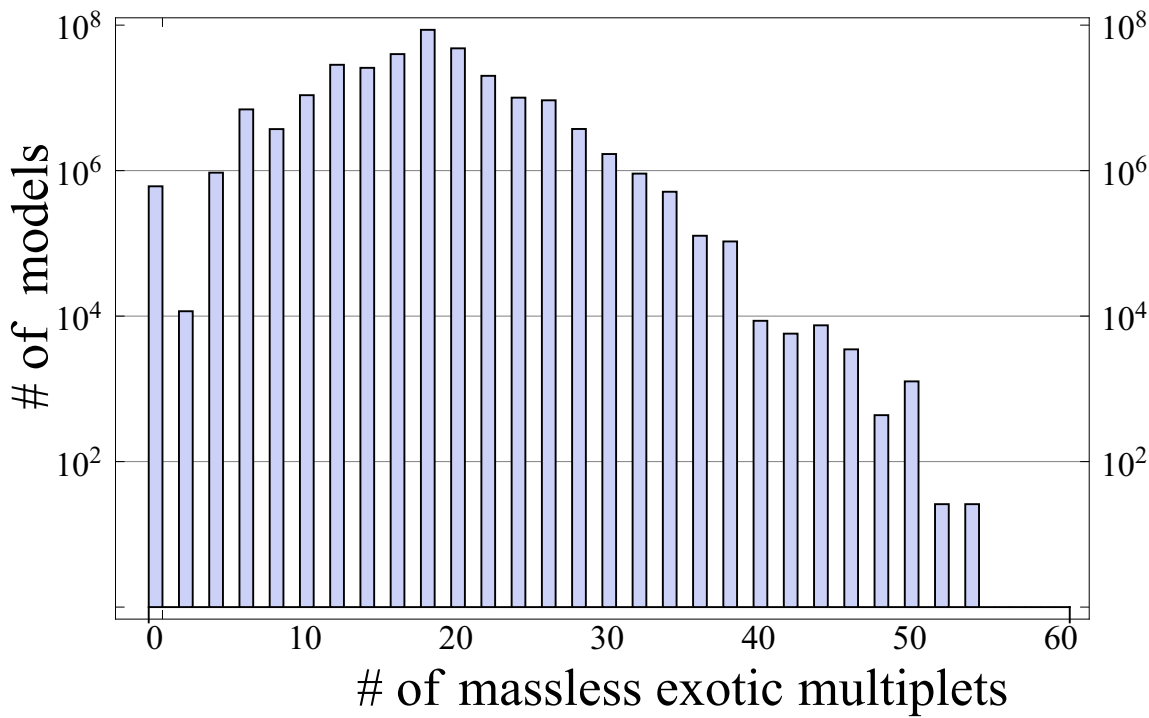}
\caption{\label{gen3_exotics}
\it Number of 3-generation models versus total number of exotic multiplets in
a random sample of $10^{11}$ GGSO configurations.}
\end{figure}

\begin{figure}[!ht]
\centering
\includegraphics[width=\textwidth]{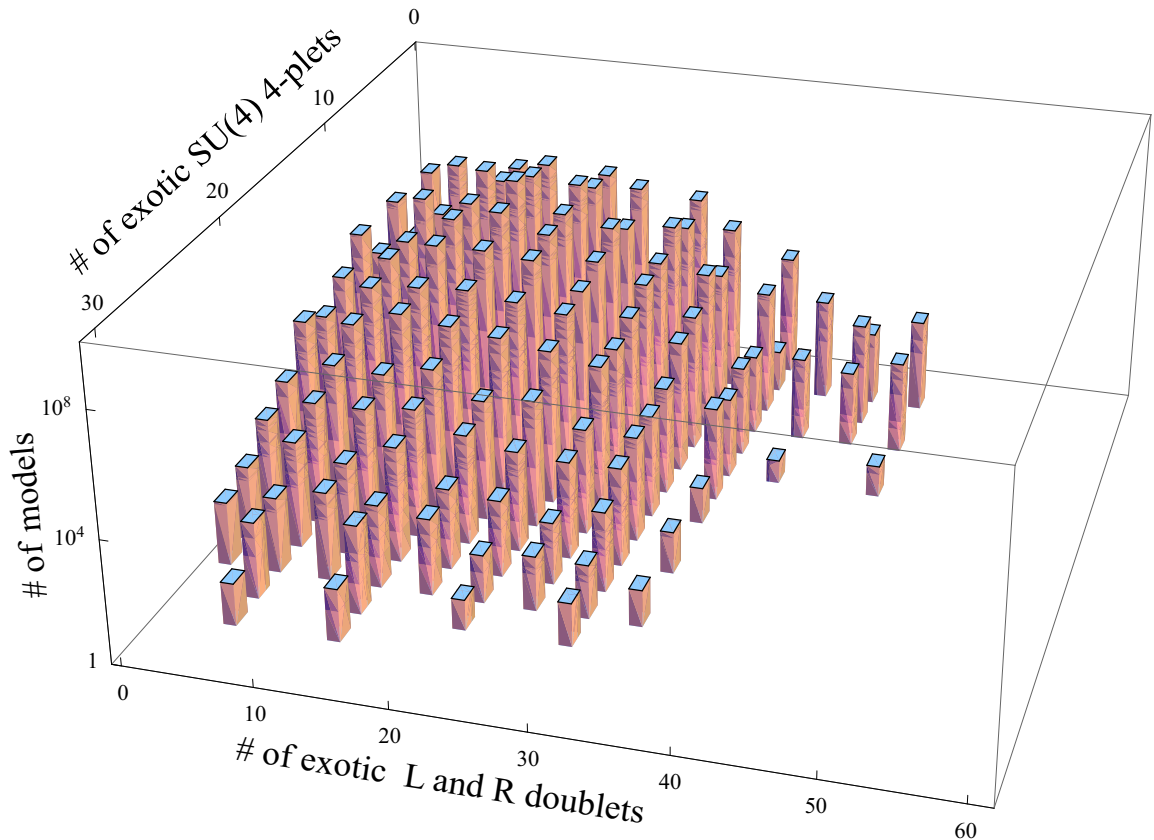}
\caption{\label{gen3_ex_det}
\it Number of 3-generation models versus  number of exotic  $SU(4)$ multiplets
and  total number of $L$ plus $R$ exotic $SU(2)$ doublets in a random sample of
$10^{11}$ GGSO configurations.
We note that the exophobic cases correspond to the upper
left column.}
\end{figure}

\begin{figure}[!ht]
\centering
\includegraphics[width=\textwidth]{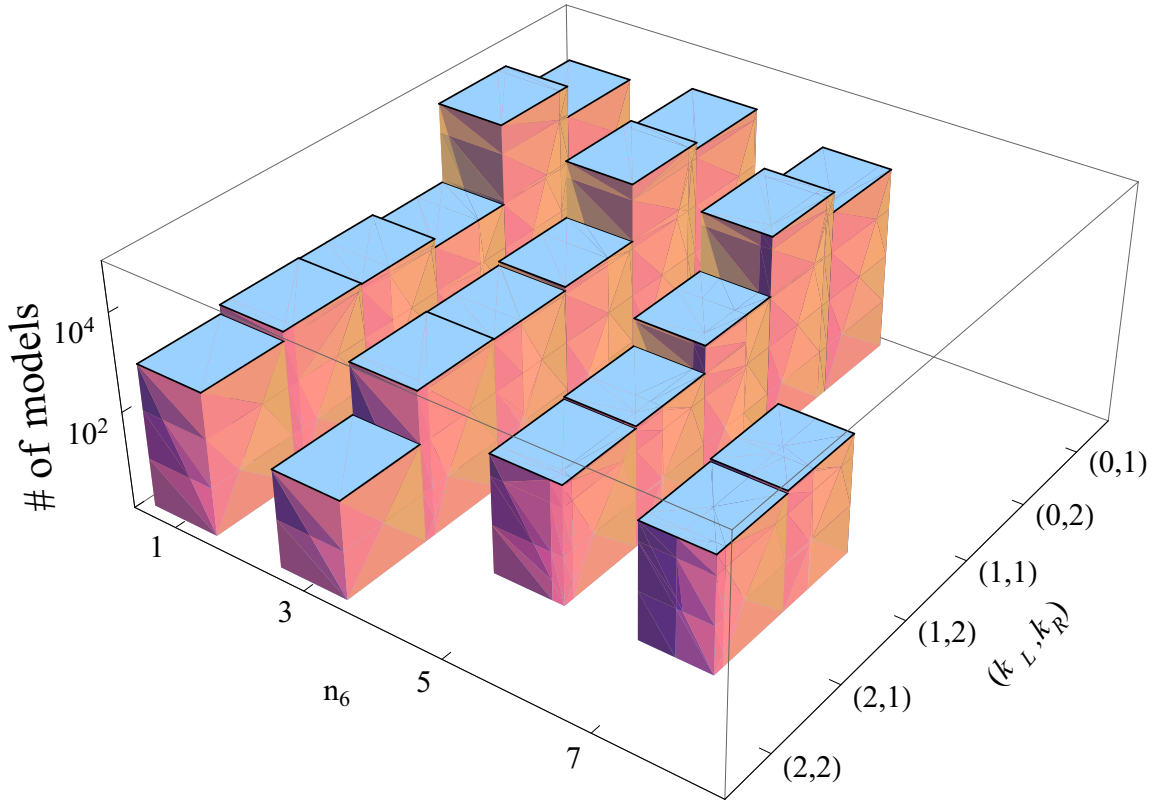}
\caption{\label{gen3_hn6_det}
\it Number of 3-generation models versus  number of additional nochiral left
and right pairs $(k_L,k_R)$ and additional   $(6,1,1)$ $SU(4)$ reps ($n_6$)  in
a random sample of $10^{11}$ GGSO configurations. We note that accommodating
the heavy Higgs states necessitates $k_R=1$. By eq. (\ref{empirical}) the
minimal case in realistic models also requires $n_6=1$.}
\end{figure}

We next explore the space of Pati--Salam free fermionic heterotic string vacua.
We perform a statistical sampling in a space of $10^{11}$ models out of the
total of $2^{51}$. Using a computer FORTRAN95 program running on a 
single node of the Theoretical Physics Division
of University of Ioannina, HPC cluster, we were able to obtain the
relative data within a period
of one week. This corresponds to examining  approximately 1:20000
models in this class.
Increasing the sample by one order of magnitude is within the
cluster capabilities, however, as
already checked by using a $10^9$ and a $10^{10}$
random sample, the results obtain are similar to the ones
presented below. Some of the results are presented in Figures
\ref{gen_distribution}-\ref{gen3_hn6_det} and Table 1.

In Figure \ref{gen_distribution} the number of models versus the number of
generations is displayed. In agreement with the results of
ref. \cite{fkr,fknr}
the number of models has a peak for models with vanishing
number of generations, and decreases
with increasing number of generations. Of note in Figure \ref{gen_distribution}
is the absence of any models with 7, 9, 11, 13, 14 and 15 generations.
This may indicate that this cases are completely forbidden or are
extremely unlikely cases in the space of all possibilities.

In Figure \ref{gen_frac} we display a three dimensional plot of
the number of models versus the number of generations and
the total number of exotic fractionally charged states. As seen from the
figure the distribution exhibits a pick for models with zero chiral generations
and a nonvanishing number of exotic multiplets, and decreases with increasing
and decreasing number of exotics. Moreover, we find no correlation between
the absence of fractionally
charge exotic states and the number of generations. We can have exophobic models
for all values of $n_g$.

However, in the case of models without any exotic
multiplets we observe the following relation between the number of chiral
generations ($n_g$), the number of Higgs
bi--doublets ($n_h$) and sextets ($n_6$)
\begin{eqnarray}
n_g \mod 2 = n_h \mod 2 = n_6 \mod 2 \label{empirical}
\end{eqnarray}
This empirical
observation is in accord with the data of the exophobic model presented in
ref. \cite{exophobic},
\noindent
and is corroborated by the data of
Table \ref{3g_nofracs} where
we display the multiplicities of models with respect to
$n_g$, $n_h$ and $n_6$. As noted from the table
the number of Higgs bi--doublets and sextets is indeed odd or even depending
on the number of generations.
Another important phenomenological point to note from table \ref{3g_nofracs}
is the existence of exophobic models with a varying number of Higgs
bi--doublets
representations. The Pati--Salam models face the potential problem of
doublet--doublet
splitting due the coupling of the Higgs bi--doublet to both the up and down
quarks,
and resulting flavor changing neutral currents transitions.
One way to alleviate the
problem is by having several Higgs bi--doublets representations, where one
gives
mass to up--type quarks and another generate masses to the down--type quarks.

\begin{table}
\centering
\begin{tabular}{|c|c|c|r|}
\hline
$n_g$&$n_h$&$n_6$&\# of models\\
\hline
 0 & 0 & 0 & 7389484 \\\hline
 0 & 0 & 2 & 1645466 \\\hline
 0 & 0 & 4 & 1000290 \\\hline
 0 & 0 & 6 & 7964 \\\hline
 0 & 0 & 8 & 35156 \\\hline
 0 & 0 & 12 & 125 \\\hline
 0 & 0 & 16 & 48 \\\hline
 0 & 2 & 0 & 1772537 \\\hline
 0 & 2 & 2 & 3370245 \\\hline
 0 & 2 & 4 & 282693 \\\hline
 0 & 2 & 6 & 101806 \\\hline
 0 & 2 & 8 & 240 \\\hline
 0 & 2 & 10 & 1425 \\\hline
 0 & 4 & 0 & 1281766 \\\hline
 0 & 4 & 2 & 314402 \\\hline
 0 & 4 & 4 & 1272994 \\\hline
 0 & 4 & 6 & 41240 \\\hline
 0 & 4 & 8 & 26600 \\\hline
 0 & 4 & 12 & 695 \\\hline
 0 & 4 & 16 & 3 \\\hline
 0 & 6 & 0 & 32801 \\\hline
 0 & 6 & 2 & 162980 \\\hline
 0 & 6 & 4 & 42929 \\\hline
 0 & 6 & 6 & 197305 \\\hline
 0 & 6 & 10 & 1077 \\\hline
 0 & 8 & 0 & 83905 \\\hline
 0 & 8 & 2 & 891 \\\hline
 0 & 8 & 4 & 44391 \\\hline
 0 & 8 & 8 & 53896 \\\hline
 0 & 8 & 10 & 667 \\\hline
 0 & 8 & 12 & 198 \\\hline
 0 & 8 & 16 & 38 \\\hline
 \end{tabular}
\
 \begin{tabular}{|c|c|c|r|}
\hline
$n_g$&$n_h$&$n_6$&\# of models\\\hline
0 & 10 & 0 & 948 \\\hline
 0 & 10 & 2 & 3951 \\\hline
 0 & 10 & 6 & 1650 \\\hline
 0 & 10 & 8 & 716 \\\hline
 0 & 10 & 10 & 2681 \\\hline
 0 & 10 & 14 & 7 \\\hline
 0 & 12 & 0 & 1657 \\\hline
 0 & 12 & 4 & 2207 \\\hline
 0 & 12 & 8 & 322 \\\hline
 0 & 12 & 12 & 2458 \\\hline
 0 & 14 & 2 & 14 \\\hline
 0 & 14 & 10 & 4 \\\hline
 0 & 16 & 0 & 336 \\\hline
 0 & 16 & 4 & 37 \\\hline
 0 & 16 & 8 & 98 \\\hline
 0 & 16 & 16 & 121 \\\hline
 0 & 18 & 2 & 3 \\\hline
 0 & 20 & 0 & 2 \\\hline
 0 & 20 & 4 & 1 \\\hline
 0 & 20 & 12 & 2 \\\hline
 0 & 24 & 0 & 2 \\\hline
 0 & 24 & 8 & 1 \\\hline
 0 & 24 & 24 & 1 \\\hline
 1 & 1 & 1 & 690074 \\\hline
 1 & 1 & 3 & 50495 \\\hline
 1 & 3 & 1 & 54719 \\\hline
 1 & 3 & 3 & 701850 \\\hline
 1 & 3 & 5 & 47239 \\\hline
 1 & 5 & 3 & 51664 \\\hline
 1 & 5 & 5 & 91419 \\\hline
 1 & 5 & 7 & 2408 \\\hline
 1 & 7 & 5 & 2636 \\\hline
 \end{tabular}
\caption{
\label{3g_nofracs}
Multiplicities  of  massless fractional
charge free models with respect to:
the number of generations $n_g$,
the number of Higgs bi--doublets $n_h$,
and the number of colour sextets $n_6$, 
in a random sample of $10^{11}$ PS models.}
\end{table}

\begin{table}
 \begin{tabular}{|c|c|c|r|}
\hline
$n_g$&$n_h$&$n_6$&\# of models\\
\hline
 1 & 7 & 7 & 2283 \\\hline
 2 & 0 & 0 & 159209 \\\hline
 2 & 0 & 4 & 2935 \\\hline
 2 & 2 & 2 & 1060873 \\\hline
 2 & 2 & 6 & 15898 \\\hline
 2 & 2 & 10 & 243 \\\hline
 2 & 4 & 0 & 4435 \\\hline
 2 & 4 & 4 & 220673 \\\hline
 2 & 4 & 8 & 1180 \\\hline
 2 & 6 & 2 & 25966 \\\hline
 2 & 6 & 6 & 53586 \\\hline
 2 & 6 & 10 & 52 \\\hline
 2 & 8 & 0 & 526 \\\hline
 2 & 8 & 4 & 1631 \\\hline
 2 & 8 & 8 & 5419 \\\hline
 2 & 10 & 2 & 824 \\\hline
 2 & 10 & 6 & 61 \\\hline
 2 & 10 & 10 & 629 \\\hline
 3 & 1 & 1 & 240224 \\\hline
 3 & 1 & 3 & 19086 \\\hline
 3 & 3 & 1 & 20709 \\\hline
 3 & 3 & 3 & 238714 \\\hline
 3 & 3 & 5 & 14007 \\\hline
 3 & 5 & 3 & 14932 \\\hline
 3 & 5 & 5 & 56886 \\\hline
 3 & 5 & 7 & 539 \\\hline
 3 & 7 & 5 & 591 \\\hline
 3 & 7 & 7 & 3135 \\\hline
 4 & 0 & 0 & 105365 \\\hline
 4 & 0 & 4 & 3234 \\\hline
  4 & 0 & 8 & 114 \\\hline
 \end{tabular}
\
 \begin{tabular}{|c|c|c|r|}
\hline
$n_g$&$n_h$&$n_6$&\# of models\\\hline
 4 & 0 & 12 & 3 \\\hline
 4 & 2 & 2 & 145699 \\\hline
 4 & 2 & 6 & 2159 \\\hline
 4 & 2 & 10 & 14 \\\hline
 4 & 4 & 0 & 4757 \\\hline
 4 & 4 & 4 & 118796 \\\hline
 4 & 4 & 8 & 1546 \\\hline
 4 & 4 & 12 & 42 \\\hline
 4 & 6 & 2 & 2660 \\\hline
 4 & 6 & 6 & 27834 \\\hline
 4 & 6 & 10 & 84 \\\hline
 4 & 8 & 0 & 556 \\\hline
 4 & 8 & 4 & 2484 \\\hline
 4 & 8 & 8 & 7942 \\\hline
 4 & 10 & 2 & 24 \\\hline
 4 & 10 & 6 & 81 \\\hline
 4 & 10 & 10 & 22 \\\hline
 4 & 12 & 0 & 37 \\\hline
 4 & 12 & 4 & 124 \\\hline
 4 & 12 & 12 & 234 \\\hline
 4 & 16 & 0 & 1 \\\hline
 5 & 1 & 1 & 5743 \\\hline
 5 & 3 & 3 & 24930 \\\hline
 5 & 5 & 5 & 16949 \\\hline
 5 & 7 & 7 & 656 \\\hline
 6 & 0 & 0 & 9339 \\\hline
 6 & 0 & 4 & 162 \\\hline
 6 & 2 & 2 & 34884 \\\hline
 6 & 2 & 6 & 55 \\\hline
 6 & 4 & 0 & 184 \\\hline
  6 & 4 & 4 & 10612 \\\hline
  \end{tabular}\\
\nolabel
Table \ref{3g_nofracs} continued.
\end{table}
\begin{table}
 \begin{tabular}{|c|c|c|r|}
 \hline
$n_g$&$n_h$&$n_6$&\# of models\\\hline
 6 & 4 & 8 & 26 \\\hline
 6 & 6 & 2 & 62 \\\hline
 6 & 6 & 6 & 7539 \\\hline
 6 & 6 & 10 & 10 \\\hline
 6 & 8 & 4 & 34 \\\hline
 6 & 8 & 8 & 781 \\\hline
 6 & 10 & 6 & 20 \\\hline
 6 & 10 & 10 & 187 \\\hline
 8 & 0 & 0 & 2543 \\\hline
 8 & 0 & 8 & 35 \\\hline
 8 & 2 & 2 & 2529 \\\hline
 8 & 4 & 4 & 7055 \\\hline
 8 & 4 & 12 & 3 \\\hline
 8 & 6 & 6 & 1742 \\\hline
 8 & 8 & 0 & 19 \\\hline
 8 & 8 & 8 & 3328 \\\hline
 8 & 8 & 16 & 1 \\\hline
 8 & 10 & 10 & 134 \\\hline
 8 & 12 & 4 & 4 \\\hline
 8 & 12 & 12 & 100 \\\hline
 8 & 16 & 8 & 3 \\\hline
 8 & 16 & 16 & 4 \\\hline
 10 & 0 & 0 & 124 \\\hline
 10 & 2 & 2 & 219 \\\hline
 10 & 4 & 4 & 112 \\\hline
 10 & 6 & 6 & 187 \\\hline
 10 & 8 & 8 & 23 \\\hline
 12 & 0 & 0 & 47 \\\hline
 12 & 2 & 2 & 22 \\\hline
 12 & 4 & 4 & 122 \\\hline
 12 & 8 & 8 & 145 \\\hline
 12 & 10 & 10 & 3 \\\hline
 12 & 12 & 12 & 43 \\\hline
 16 & 0 & 0 & 7 \\\hline
 16 & 4 & 4 & 17 \\\hline
 16 & 8 & 8 & 7 \\\hline
 16 & 12 & 12 & 4\\\hline
\end{tabular}\\
\nolabel
Table \ref{3g_nofracs} continued.
\end{table}

In figure \ref{gen_no_fracs} we display the multiplicities of models versus the
number of generations in the case of exotic free models. As seen from the
figure
the number of models decreases with increasing number of generations. The same
exclusion of models with some number of generations noted in figure
\ref{gen_distribution} is also seen in figure \ref{gen_frac} for the
same cases.

Figure \ref{gen3_exotics} displays the total number of three generation models
versus
the number of exotic fractionally charged states in a given three generation
model.
As seen from the figure the total number of exophobic three generation models
is
slightly less than $10^6$, which is roughly $1/10^5$ from the entire sample.
Hence we can surmise that exophobia is a common feature in the sampled space
of string vacua. Having established a quasi--realistic spectrum the next stage
is to analyse the Yukawa couplings in the models.
The abundance of exotic free three generation models suggests that
models with viable Yukawa and fermion mass spectrum do exist in this space of
string vacua.

\newpage

In figure \ref{gen3_ex_det} we display in a three dimensional plot
the total number of three generation models versus the number of
exotic $SU(4)$ \textbf{4}--plets and number of exotic $SO(4)$
$\textbf{2}_\textbf{L}$ and $\textbf{2}_\textbf{R}$
doublets. In figure \ref{gen3_hn6_det} we display in a three dimensional
plot the number of three generation models versus the number of
additional non--chiral representations in the 
${({\bar{\textbf{4}}},\textbf{1},\textbf{2}_\textbf{R})\oplus(\textbf{4},
\textbf{1},\textbf{2}_\textbf{R})}$
and
${({\textbf{4}},\textbf{2}_\textbf{L},\textbf{1}})
\oplus({\bar{\textbf{4}}},\textbf{2}_{\textbf{L}},\textbf{1})$
and additional $(\textbf{6},\textbf{1},\textbf{1})$ multiplets
of $SU(4)\times SU(2)_L\times SU(2)_R$. Finally in table \ref{summary}
we tabulate the number of models with sequential imposition of phenomenological
constraints. The total number of models in the sample is $10^{11}$. We first
impose that there is no enhancement of the four dimensional gauge symmetry.
Roughly
80\% percent of the models satisfy this criteria. Next we impose that the
generations form complete families. That is there is no chiral representation
of the Pati--Salam gauge group that is not accompanied by either the
representation that
completes it to a representation of $SO(10)$ or renders it non--chiral. So the
entire
chiral spectrum is contained in complete representations of $SO(10)$ decomposed
under the Pati--Salam subgroup. Roughly 1/5 of the previous set satisfy this
criterion.
The restriction to three chiral generations reduces further the number of
models by two
orders of magnitude. Imposing the existence of heavy string states to break the
Pati--Salam gauge symmetry to the Standard Model gauge group leads to a
reduction by
another order of magnitude. The requirement of Standard Model Higgs doublets
does not lead to a further reduction because as noted above in eq.
\eqref{empirical}
the total number of Higgs bi--doublets is equal to the number of chiral
generations modulo 2. Therefore, existence of three chiral generations
necessarily
implies a non--zero number of Higgs bi--doublets to be in the spectrum.
Finally,
imposing the absence of massless exotics reduces the number of models
by further two orders of magnitude. Therefore, the reduction from the initial
sample is by roughly six order of magnitude, {\it i.e.} one in every $10^6$
models
satisfy all of these constraints. Given that the total number of vacua in the
space
of models scanned is of the order of $10^{15}$, we expect that $10^9$ of the
models
satisfy these criteria, which leaves a substantial number to accommodate
further
phenomenological constraints. For example, requiring minimal number of PS breaking Higgs
($k_L=0,k_R=1$) truncates further by 4 the number of models as seen in line (g).
Furthermore, approximately 1/4 of these models have also minimal
Standard Model Higgs sector with ($n_h=1$)

\begin{table}
\begin{tabular}{|c|l|r|c|c|r|}
\hline
&constraint & \parbox[c]{2.5cm}{\# of models\\ in sample }& probability
&\parbox[c]{3cm}{ estimated \# of \\ models in class}\\
\hline
 & None & $100000000000$ & $1$ &$2.25\times 10^{15}$ \\ \hline
(a)&\parbox[c][1cm]{4.5cm}{+ No gauge group\\ enhancements.} & 78977078333 &
$7.90\times10^{-1}$ &   $1.78\times 10^{15}$ \\  \hline
(b)&{+ Complete families} & 22497003372 & $2.25\times 10^{-1}$ & $5.07\times
10^{14}$ \\  \hline
(c)&{+ 3 generations } & 298140621 & $2.98 \times 10^{-3}$ & $6.71\times
10^{12}$ \\  \hline
(d)&{+ PS breaking Higgs } & 23694017 & $2.37 \times 10^{-4}$ & $5.34\times
10^{11}$ \\\hline
(e)&{+ SM breaking Higgs } & 19191088 & $1.92 \times 10^{-4}$ & $4.32\times
10^{11}$ \\\hline
(f)&{+ No massless exotics } & 121669 & $1.22 \times 10^{-6}$ & $2.74\times
10^9$ \\\hline
(g)&{+ Minimal PS Higgs } & 31804 & $3.18 \times 10^{-7}$ & $7.16\times
10^8$ \\\hline
\end{tabular}
\caption{
\label{summary}
Pati--Salam models statistics with respect to phenomenological constraints
imposed on massless spectrum. Constraints in second column
act additionally. Omitting constraint (e) does not change the results of (f), (g)
since all massless exotic free models have an odd number of pairs of SM Higgs doublets. }
\end{table}

\newpage

\section{Conclusions}

The Standard Model data supports the embedding of its matter spectrum
into spinorial 16 representations of $SO(10)$. Indeed, the
augmentation of the Standard Model by the right--handed neutrinos,
proposed originally by Pati and Salam \cite{ps},
was corroborated by terrestrial and astrophysical neutrino experiments.
String theory enables the construction of phenomenological
models that provide the arena to explore the
synthesis of gravity and the gauge interaction within a self--consistent
framework. It is desirable that such phenomenological string models
preserve the $SO(10)$ embedding of the Standard Model matter states,
while its Higgs representations are obtained from the vectorial 10
representation.

Absence of adjoint Higgs representations in models with level one
Kac--Moody algebras necessitates that the $SO(10)$ symmetry is broken
directly at the string level. Heterotic string models in the free fermionic
formulation produce such three generation models that preserve the
$SO(10)$ embedding of the Standard Model spectrum.
Early constructions of such models, constructed in the late eighties, consisted
of isolated examples. During the last few years
systematic methods to classify large classes of symmetric free fermionic
models were developed. Initially these methods were applied to the
classification of models with unbroken $SO(10)$ GUT symmetry,
with respect to the number of generations, \ie of the
difference between spinorial and anti--spinorial representations,
and subsequently also with respect to vectorial representations.
The classification revealed a new duality symmetry in the space
of vacua under exchange of spinor and vector representations.

In this paper we extended the classification to models in which the
$SO(10)$ GUT symmetry is broken to the Pati--Salam subgroup.
A generic feature of such string models in which the $SO(10)$ symmetry is
broken and that preserve the canonical GUT embedding of the weak
hypercharge, is the appearance of exotic fractionally charged states
in the string spectrum. Such states are severely constrained by
experimental observations. The reason being that the lightest of these
states is stable due to electric charge conservation,
and must be sufficiently massive and diluted in a viable model.
One possibility is that the harmful states only exist in the massive
string spectrum. In ref. \cite{exophobic} we presented an
explicit example of such an exophobic
quasi--realistic Pati--Salam heterotic--string model.
It is of interest to study whether such exophobic string models
are also obtained in other classes of orbifold models \cite{raby}.
We also note that, provided that they satisfy all the
observational constraints, the exotic states may produce
stable string relics \cite{fc} that are of further interest.

Furthermore, we elaborated on the classification method that enabled
the discovery of the exophobic model in \cite{exophobic}.
The key to obtaining this result is the extension of the
algebraic expressions derived in refs. \cite{fkr} for spinorial
and vectorial $SO(10)$ representations to all the sectors
in the string models. This enables the derivation of algebraic
formulas for the entire spectrum that arises in the string models.
These formulas are used in a computer code, and enables us to
scan a space of $2^{51}$ models. This number of vacua is too large for a
complete classification and we performed a statistical analysis
that samples $10^{11}$ models in this class of vacua.
Imposing various phenomenological criteria we find that roughly one in
$10^6$ of the models pass similar phenomenological impositions as
the exophobic model of ref. \cite{exophobic}. This suggests that sufficient
freedom remains in the space of vacua to satisfy the additional constraints
required by the Standard Model data.

Having at our disposal a plethora of semi--realistic $N=1$ string
vacua with the full massless and massive spectrum give us the
possibility to study not only their phenomenological properties,
but also their cosmological implications, once supersymmetry
breaking is incorporated. Following the lines of refs. \cite{cosmological}
the cosmological evolution of all these models can be studied since
the exact one--loop free energy and effective potential
can be calculated at the string level, at least for models in which
supersymmetry breaking is achieved via geometrical fluxes
\cite{cosmological}. This will lead
to a cosmological evolutionary behavior at least for temperature below the
Hagedron era and before the electroweak phase transition, thanks to the
attractor mechanism valid in this intermediate cosmological regime
\cite{cosmological}.

Another direction along these lines is to check the possible
deformations induced by the moduli participating in the
supersymmetry breaking \cite{msdsdefo}, and to select the low energy vacua,
which lead to Hagedron and initial singularity free models
at early cosmological times \cite{msdsdefo}.

Finally, after the electroweak phase transitions, one can derive
in full generality the soft supersymmetry breaking terms \cite{zwirner}
in the low effective field theory, once the SUSY breaking
fluxes (geometrical or not) are suitably fixed.

\section{Acknowledgements}

AEF and JR would like to thank the Ecole Normal Sup\'erieur, and BA, CK and JR
would like to thank the University of Liverpool, for hospitality.
AEF is supported in part by STFC under contract PP/D000416/1.
CK is supported in part by the EU under the contracts
MRTN-CT-2004-005104, MRTN-CT-2004-512194,
MRTN-CT-2004-503369, MEXT-CT-2003-509661, 
CNRS PICS 2530, 3059 and 3747, 
ANR (CNRS-USAR) contract 05-BLAN-0079-01 and 
ANR programme blanc NT09-573739, and by the IFCPAR programme 4104-2.
JR work is supported in part by the EU under contracts
MRTN--CT--2006--035863--1, MRTN--CT--2004--503369 and
PITN--GA--2009--237920.

\appendix

\section{Hidden matter states,representations and projectors}\label{appendixa}

The expressions for the projectors corresponding to $B_{pqrs}^{(4,5,6)}$ from
\eqref{eqn:hidspin1} are given below
\begin{eqnarray}
P_{pqrs}^{(4)}&=&
\frac{1}{8}\,\left(1-c\binom{e_1}{B_{pqrs}^{(4)}} \right)\,
\cdot\left(1-c\binom{e_2}{B_{pqrs}^{(4)}}\right)\,  \nonumber\ \\
&&\cdot\left(1-c\binom{z_2}{B_{pqrs}^{(4)}}\right) \nonumber\\
P_{pqrs}^{(5)}&=&
\frac{1}{8}\,\left(1-c\binom{e_3}{B_{pqrs}^{(5)}}\right)\,
\cdot\left(1-c\binom{e_4}{B_{pqrs}^{(5)}}\right)\,  \nonumber\\\
&&\cdot\left(1-c\binom{z_2}{B_{pqrs}^{(5)}}\right) \\
P_{pqrs}^{(6)}&=&
\frac{1}{8}\,\left(1-c\binom{e_5}{B_{pqrs}^{(6)}}\right)\,
\cdot\left(1-c\binom{e_6}{B_{pqrs}^{(6)}}\right)\,  \nonumber\\\
&&\cdot\left(1-c\binom{z_2}{B_{pqrs}^{(6)}}\right) \nonumber
\end{eqnarray}
Their corresponding analytic expressions are
\ba
\left(
\begin{array}{cccc}
\oo{e_1}{e_3}&\oo{e_1}{e_4}&\oo{e_1}{e_5}&\oo{e_1}{e_6}\\
\oo{e_2}{e_3}&\oo{e_2}{e_4}&\oo{e_2}{e_5}&\oo{e_2}{e_6}\\
\oo{z_2}{e_3}&\oo{z_2}{e_4}&\oo{z_2}{e_5}&\oo{z_2}{e_6}
\end{array}
\right)
\left(
\begin{array}{c}
p\\
q\\
r\\
s\\
\end{array}
\right)
=
\left(
\begin{array}{c}
\oo{e_1}{b_1+ x + z_{1}}\\
\oo{e_2}{b_1+ x + z_{1}}\\
\oo{z_2}{b_1+ x + z_{1}}
\end{array}
\right) \nonumber
\ea

\vspace{-0.4cm}

\ba
\left(
\begin{array}{cccc}
\oo{e_3}{e_1}&\oo{e_3}{e_2}&\oo{e_3}{e_5}&\oo{e_3}{e_6}\\
\oo{e_4}{e_1}&\oo{e_4}{e_2}&\oo{e_4}{e_5}&\oo{e_4}{e_6}\\
\oo{z_2}{e_1}&\oo{z_2}{e_2}&\oo{z_2}{e_5}&\oo{z_2}{e_6}
\end{array}
\right)
\left(
\begin{array}{c}
p\\
q\\
r\\
s\\
\end{array}
\right)
=
\left(
\begin{array}{c}
\oo{e_1}{b_2+ x + z_{1}}\\
\oo{e_2}{b_2+ x + z_{1}}\\
\oo{z_2}{b_2+ x + z_{1}}
\end{array}
\right)
\ea

\vspace{-0.4cm}

\ba
\left(
\begin{array}{cccc}
\oo{e_5}{e_1}&\oo{e_5}{e_2}&\oo{e_5}{e_3}&\oo{e_5}{e_4}\\
\oo{e_6}{e_1}&\oo{e_6}{e_2}&\oo{e_6}{e_3}&\oo{e_6}{e_4}\\
\oo{z_2}{e_1}&\oo{z_2}{e_2}&\oo{z_2}{e_3}&\oo{z_2}{e_4}
\end{array}
\right)
\left(
\begin{array}{c}
p\\
q\\
r\\
s\\
\end{array}
\right)
=
\left(
\begin{array}{c}
\oo{e_1}{b_3+ x + z_{1}}\\
\oo{e_2}{b_3+ x + z_{1}}\\
\oo{z_2}{b_3+ x + z_{1}}
\end{array}
\right) \nonumber
\ea
The remaining 48 projectors corresponding to hidden sectors given in
$(\ref{eqn:hidspin2})$ are given by
\begin{eqnarray}
P_{pqrs}^{(7)}&=&
\frac{1}{4}\,\left(1-c\binom{e_1}{B_{pqrs}^{(7)}} \right)\,\nonumber
\cdot\left(1-c\binom{e_2}{B_{pqrs}^{(7)}}\right)\, \nonumber\\
&&\cdot\frac{1}{4}\left(1-c\binom{z_1}{B_{pqrs}^{(7)}}\right)\,\nonumber
\cdot\left(1-c\binom{\alpha}{B_{pqrs}^{(7)}}\right)\, \nonumber\\
P_{pqrs}^{(8)}&=&
\frac{1}{4}\,\left(1-c\binom{e_3}{B_{pqrs}^{(8)}}\right)\, \nonumber
\cdot\left(1-c\binom{e_4}{B_{pqrs}^{(8)}}\right)\, \\
&&\cdot\frac{1}{4}\left(1-c\binom{z_1}{B_{pqrs}^{(8)}}\right)\,
\cdot\left(1-c\binom{\alpha}{B_{pqrs}^{(8)}}\right)\\
P_{pqrs}^{(9)}&=&
\frac{1}{4}\,\left(1-c\binom{e_5}{B_{pqrs}^{(9)}}\right)\,\nonumber
\cdot\left(1-c\binom{e_6}{B_{pqrs}^{(9)}}\right)\, \nonumber\\
&&\cdot\frac{1}{4}\left(1-c\binom{z_1}{B_{pqrs}^{(9)}}\right) \nonumber
\cdot\left(1-c\binom{\alpha}{B_{pqrs}^{(9)}}\right)\,\nonumber
\end{eqnarray}
The analytic expressions for $P_{p,q,r,s}^{7,8,9}$ are given below:
\ba
\left(
\begin{array}{cccc}
\oo{e_1}{e_3}&\oo{e_1}{e_4}&\oo{e_1}{e_5}&\oo{e_1}{e_6}\\
\oo{e_2}{e_3}&\oo{e_2}{e_4}&\oo{e_2}{e_5}&\oo{e_2}{e_6}\\
\oo{z_1}{e_3}&\oo{z_1}{e_4}&\oo{z_1}{e_5}&\oo{z_1}{e_6}\\
\oo{\alpha}{e_3}&\oo{\alpha}{e_4}&\oo{\alpha}{e_5}&\oo{\alpha}{e_6}\\
\end{array}
\right)
\left(
\begin{array}{c}
p\\
q\\
r\\
s\\
\end{array}
\right)
=
\left(
\begin{array}{c}
\oo{e_1}{b_1+ x + z_{2}}\\
\oo{e_2}{b_1+ x + z_{2}}\\
\oo{z_1}{b_1+ x + z_{2}}\\
\oo{\alpha}{b_1+ x + z_{2}}
\end{array}
\right) \nonumber
\ea

\vspace{-0.2cm}

\ba
\left(
\begin{array}{cccc}
\oo{e_3}{e_1}&\oo{e_3}{e_2}&\oo{e_3}{e_5}&\oo{e_3}{e_6}\\
\oo{e_4}{e_1}&\oo{e_4}{e_2}&\oo{e_4}{e_5}&\oo{e_4}{e_6}\\
\oo{z_1}{e_1}&\oo{z_1}{e_2}&\oo{z_1}{e_5}&\oo{z_1}{e_6}\\
\oo{\alpha}{e_1}&\oo{\alpha}{e_2}&\oo{\alpha}{e_5}&\oo{\alpha}{e_6}\\
\end{array}
\right)
\left(
\begin{array}{c}
p\\
q\\
r\\
s\\
\end{array}
\right)
=
\left(
\begin{array}{c}
\oo{e_1}{b_2+ x + z_{2}}\\
\oo{e_2}{b_2+ x + z_{2}}\\
\oo{z_1}{b_2+ x + z_{2}}\\
\oo{\alpha}{b_2+ x + z_{2}}
\end{array}
\right)
\ea

\vspace{-0.2cm}

\ba
\left(
\begin{array}{cccc}
\oo{e_5}{e_1}&\oo{e_5}{e_2}&\oo{e_5}{e_3}&\oo{e_5}{e_4}\\
\oo{e_6}{e_1}&\oo{e_6}{e_2}&\oo{e_6}{e_3}&\oo{e_6}{e_4}\\
\oo{z_1}{e_1}&\oo{z_1}{e_2}&\oo{z_1}{e_3}&\oo{z_1}{e_4}\\
\oo{\alpha}{e_1}&\oo{\alpha}{e_2}&\oo{\alpha}{e_3}&\oo{\alpha}{e_4}\\
\end{array}
\right)
\left(
\begin{array}{c}
p\\
q\\
r\\
s\\
\end{array}
\right)
=
\left(
\begin{array}{c}
\oo{e_1}{b_3+ x + z_{2}}\\
\oo{e_2}{b_3+ x + z_{2}}\\
\oo{z_1}{b_3+ x + z_{2}}\\
\oo{\alpha}{b_3+ x + z_{2}}
\end{array}
\right) \nonumber
\ea

\subsection{Exotic states, representations and projectors}

The representations and observable charges of   $B_{p,q,r,s}^{10,11,12}$ in
$(\ref{eqn:exospin1})$ and  $B_{p,q,r,s}^{13,14,15}$ are given below :\\
\\
\begin{tabular}{c|c|c|c|c}
 representation & $\bar{\psi}^{1,2,3}$ & $\bar{\phi}^{1,2}$ or
$\bar{\phi}^{3,4}$ & $Y$ & $Q_{em}$ \\
\hline
& $(+,+,+)$ & $(+,+)$ & 1/2& 1/2\\
$(\bar{\mathbf{4}},\mathbf{1},\mathbf{2})$ & $(+,+,+)$ & $(-,-)$ & 1/2& 1/2\\
& ($+,-,-$)& $(+,+)$ & -1/6& -1/6\\
& ($+,-,-$)& $(-,-)$ & -1/6& -1/6\\
\hline
\hline
& $(-,-,-)$ & $(-,-)$ & -1/2& -1/2\\
$(\mathbf{4},\mathbf{1},\mathbf{2})$ & $(-,-,-)$ & $(+,+)$ & -1/2& -1/2\\
& ($+,+,-$)& $(-,-)$ & 1/6& 1/6\\
& ($+,+,-$)& $(+,+)$ & 1/6& 1/6\\
\hline
\hline
$(\bar{\mathbf{4}},\mathbf{2},\mathbf{1})$ & $(+,+,+)$ & ($+,-$)& 1/2& 1/2\\
& ($+,-,-$)& ($+,-$)& -1/6& -1/6\\
\hline
\hline
$(\mathbf{4},\mathbf{2},\mathbf{1})$ & $(-,-,-)$ & ($+,-$)& -1/2& -1/2\\
& ($+,+,-$)& ($+,-$)& 1/6& 1/6\\
\end{tabular}\\

We can therefore summarise all the previous results by saying that sectors
coming from  $B_{p,q,r,s}^{10,11,12,13,14,15}$,  give rise to
$(\mathbf{4},1,1)$ and $(\mathbf{\bar{4}},1,1)$ representations under the S.M
gauge group, with fractional electric charges: $\pm{\frac{1}{2}}$ and
$\pm{\frac{1}{6}}$. \\The projectors corresponding to  $B_{p,q,r,s}^{10,11,12}$
 are:
\begin{eqnarray}
P_{pqrs}^{(10)}&=&
\frac{1}{4}\,\left(1-c\binom{e_1}{B_{pqrs}^{(10)}} \right)\,
\cdot\left(1-c\binom{e_2}{B_{pqrs}^{(10)}}\right)\, \nonumber\\
&&\cdot\frac{1}{4}\left(1-c\binom{z_2}{B_{pqrs}^{(10)}}\right)\,
\cdot\left(1-c\binom{\alpha + z_1}{B_{pqrs}^{(10)}}\right) \nonumber\\
P_{pqrs}^{(11)}&=&
\frac{1}{4}\,\left(1-c\binom{e_3}{B_{pqrs}^{(11)}} \right)\,
\cdot\left(1-c\binom{e_4}{B_{pqrs}^{(11)}}\right)\,  \nonumber\\
&&\cdot\frac{1}{4}\left(1-c\binom{z_2}{B_{pqrs}^{(11)}}\right)\,
\cdot\left(1-c\binom{\alpha + z_1}{B_{pqrs}^{(11)}}\right) \\
P_{pqrs}^{(12)}&=&
\frac{1}{4}\,\left(1-c\binom{e_5}{B_{pqrs}^{(12)}} \right)\,
\cdot\left(1-c\binom{e_6}{B_{pqrs}^{(12)}}\right)\,  \nonumber\\
&&\cdot\frac{1}{4}\left(1-c\binom{z_2}{B_{pqrs}^{(12)}}\right)\,
\cdot\left(1-c\binom{\alpha + z_1}{B_{pqrs}^{(12)}}\right) \nonumber
\end{eqnarray}
 We can get the expressions for $P^{13,14,15}$ if we substitute $B^{10,11,12}
\rightarrow B^{13,14,15}$ and $\alpha + z_{1} \rightarrow \alpha$.\\The matrix
formalism for the previous expressions is:

\ba
\left(
\begin{array}{cccc}
\oo{e_1}{e_3}&\oo{e_1}{e_4}&\oo{e_1}{e_5}&\oo{e_1}{e_6}\\
\oo{e_2}{e_3}&\oo{e_2}{e_4}&\oo{e_2}{e_5}&\oo{e_2}{e_6}\\
\oo{z_2}{e_3}&\oo{z_2}{e_4}&\oo{z_2}{e_5}&\oo{z_2}{e_6}\\
\oo{\alpha + z_{1}}{e_3}&\oo{\alpha + z_1}{e_4}&\oo{\alpha +
z_1}{e_5}&\oo{\alpha + z_1}{e_6}
\end{array}
\right)
\left(
\begin{array}{c}
p\\
q\\
r\\
s\\
\end{array}
\right)
=
\left(
\begin{array}{c}
\oo{e_1}{b_1 +  \alpha}\\
\oo{e_2}{b_1 + \alpha}\\
\oo{z_1}{b_1 + \alpha}\\
\oo{z_2}{b_1 + \alpha}
\end{array}
\right) \nonumber
\ea

\vspace{-0.2cm}

\ba
\left(
\begin{array}{cccc}
\oo{e_3}{e_1}&\oo{e_3}{e_2}&\oo{e_3}{e_5}&\oo{e_3}{e_6}\\
\oo{e_4}{e_1}&\oo{e_4}{e_2}&\oo{e_4}{e_5}&\oo{e_4}{e_6}\\
\oo{z_2}{e_1}&\oo{z_2}{e_2}&\oo{z_2}{e_5}&\oo{z_2}{e_6}\\
\oo{\alpha + z_{1}}{e_1}&\oo{\alpha + z_1}{e_2}&\oo{\alpha +
z_1}{e_5}&\oo{\alpha + z_1}{e_6}
\end{array}
\right)
\left(
\begin{array}{c}
p\\
q\\
r\\
s\\
\end{array}
\right)
=
\left(
\begin{array}{c}
\oo{e_1}{b_2 +  \alpha}\\
\oo{e_2}{b_2 + \alpha}\\
\oo{z_1}{b_2 + \alpha}\\
\oo{z_2}{b_2 + \alpha}
\end{array}
\right)
\ea

\vspace{-0.2cm}

\ba
\left(
\begin{array}{cccc}
\oo{e_5}{e_1}&\oo{e_5}{e_2}&\oo{e_5}{e_3}&\oo{e_5}{e_4}\\
\oo{e_6}{e_1}&\oo{e_6}{e_2}&\oo{e_6}{e_3}&\oo{e_6}{e_4}\\
\oo{z_2}{e_1}&\oo{z_2}{e_2}&\oo{z_2}{e_3}&\oo{z_2}{e_4}\\
\oo{\alpha + z_{1}}{e_1}&\oo{\alpha + z_1}{e_2}&\oo{\alpha +
z_1}{e_3}&\oo{\alpha + z_1}{e_4}
\end{array}
\right)
\left(
\begin{array}{c}
p\\
q\\
r\\
s\\
\end{array}
\right)
=
\left(
\begin{array}{c}
\oo{e_1}{b_3 +  \alpha}\\
\oo{e_2}{b_3 + \alpha}\\
\oo{z_1}{b_3 + \alpha}\\
\oo{z_2}{b_3 + \alpha}
\end{array}
\right) \nonumber
\ea

We can get the analytical expressions for $P^{13,14,15}$ if we substitute
$\alpha + z_1 \rightarrow \alpha$.

The representations and observable charges of   $B_{p,q,r,s}^{16,17,18} + z_1$
in $(\ref{eqn:exospin2})$ and  $B_{p,q,r,s}^{19,20,21}$ are given below:\\
\\
\begin{tabular}{c|c|c|c|c}
 representation & $\bar{\psi}^{4,5}$ & $\bar{\phi}^{1,2}$ or $\bar{\phi}^{3,4}$
& $Y$ & $Q_{em}$ \\
\hline
& $(+,+)$ & $(+,+)$ & 1/2& 1/2\\
$((\mathbf{1},\mathbf{2}),(\mathbf{1},\mathbf{2}))$ & $(+,+)$ & $(-,-)$ & 1/2&
1/2\\
& ($-,-$)& $(+,+)$ & -1/2& -1/2\\
& ($-,-$)& $(-,-)$ & -1/2& -1/2\\
\hline
\hline
$((\mathbf{1},\mathbf{2}),(\mathbf{2},\mathbf{1}))$ & $(+,+)$ & $(+,-)$ & 1/2&
1/2\\
& ($-,-$)& $(+,-)$ & -1/2& -1/2\\
\hline
\hline
$((\mathbf{2},\mathbf{1}),(\mathbf{1},\mathbf{2}))$ & $(+,-)$ & ($+,+$)& 0&
-1/2,1/2\\
& ($+,-$)& ($-,-$)& 0& -1/2,1/2\\
\hline
\hline
$((\mathbf{2},\mathbf{1}),(\mathbf{2},\mathbf{1}))$ & $(+,-)$ & ($+,-$)& 0&
-1/2,1/2\\
\end{tabular}\\

The mixed states from $B_{p,q,r,s}^{16,17,18,19,20,21}$  give rise to
$(\mathbf{1,2,1})$ and $(\mathbf{1,1,2})$ representations under the Standard
Model gauge group with fractional electric charges: $\pm{\frac{1}{2}}$. The
projectors corresponding to  $B_{p,q,r,s}^{16,17,18}$ are:
\begin{eqnarray}
P_{pqrs}^{(16)}&=&
\frac{1}{8}\,\left(1-c\binom{e_1}{B_{pqrs}^{(16)}} \right)\,
\cdot\left(1-c\binom{e_2}{B_{pqrs}^{(16)}}\right)\,
\cdot\left(1-c\binom{z_2}{B_{pqrs}^{(16)}}\right)  \nonumber
\\
P_{pqrs}^{(17)}&=&
\frac{1}{8}\,\left(1-c\binom{e_3}{B_{pqrs}^{(17)}} \right)\,
\cdot\left(1-c\binom{e_4}{B_{pqrs}^{(17)}}\right)\,
\cdot\left(1-c\binom{z_2}{B_{pqrs}^{(17)}}\right)
\\
P_{pqrs}^{(18)}&=&
\frac{1}{8}\,\left(1-c\binom{e_5}{B_{pqrs}^{(18)}} \right)\,
\cdot\left(1-c\binom{e_6}{B_{pqrs}^{(18)}}\right)\,
\cdot\left(1-c\binom{z_2}{B_{pqrs}^{(18)}}\right)  \nonumber
\end{eqnarray}
In order to get the expressions for  $P_{p,q,r,s}^{19,20,21}$ we have to
substitute  $B_{p,q,r,s}^{16,17,18} \rightarrow  B_{p,q,r,s}^{19,20,21} $.
\ba
\left(
\begin{array}{cccc}
\oo{e_1}{e_3}&\oo{e_1}{e_4}&\oo{e_1}{e_5}&\oo{e_1}{e_6}\\
\oo{e_2}{e_3}&\oo{e_2}{e_4}&\oo{e_2}{e_5}&\oo{e_2}{e_6}\\
\oo{z_2}{e_3}&\oo{z_2}{e_4}&\oo{z_2}{e_5}&\oo{z_2}{e_6}\\
\end{array}
\right)
\left(
\begin{array}{c}
p\\
q\\
r\\
s\\
\end{array}
\right)
=
\left(
\begin{array}{c}
\oo{e_1}{b_1 +  \alpha + x}\\
\oo{e_2}{b_1 + \alpha +  x}\\
\oo{z_2}{b_1 + \alpha +  x}
\end{array}
\right)   \nonumber
\ea

\vspace{-0.3cm}

\ba
\left(
\begin{array}{cccc}
\oo{e_3}{e_1}&\oo{e_3}{e_2}&\oo{e_3}{e_5}&\oo{e_3}{e_6}\\
\oo{e_4}{e_1}&\oo{e_4}{e_2}&\oo{e_4}{e_5}&\oo{e_4}{e_6}\\
\oo{z_2}{e_1}&\oo{z_2}{e_2}&\oo{z_2}{e_5}&\oo{z_2}{e_6}\\

\end{array}
\right)
\left(
\begin{array}{c}
p\\
q\\
r\\
s\\
\end{array}
\right)
=
\left(
\begin{array}{c}
\oo{e_1}{b_2 +  \alpha + x}\\
\oo{e_2}{b_2 + \alpha +  x}\\
\oo{z_2}{b_2 + \alpha +  x}
\end{array}
\right)
\ea

\vspace{-0.3cm}

\ba
\left(
\begin{array}{cccc}
\oo{e_5}{e_1}&\oo{e_5}{e_2}&\oo{e_5}{e_3}&\oo{e_5}{e_4}\\
\oo{e_6}{e_1}&\oo{e_6}{e_2}&\oo{e_6}{e_3}&\oo{e_6}{e_4}\\
\oo{z_2}{e_1}&\oo{z_2}{e_2}&\oo{z_2}{e_3}&\oo{z_2}{e_4}\\
\end{array}
\right)
\left(
\begin{array}{c}
p\\
q\\
r\\
s\\
\end{array}
\right)
=
\left(
\begin{array}{c}
\oo{e_1}{b_3 +  \alpha + x}\\
\oo{e_2}{b_3 + \alpha +  x}\\
\oo{z_2}{b_3 + \alpha +  x}
\end{array}
\right)  \nonumber
\ea

We can get the analytical expressions for
$P^{19,20,21}$ if we substitute
$\alpha + x \rightarrow \alpha + x + z_1$\\

\subsection{Vectorial states, representations and projectors}

The corresponding projectors to the vectorial representations of
$(\ref{eqn:vecto})$ are:
\begin{eqnarray}
P_{pqrs}^{(i)(\bar{\psi}_{123})}&=&
\frac{1}{4}\,\left(1-c\binom{e_{2i-1}}{B_{pqrs}^{(i)}+x} \right)\,
\cdot\left(1-c\binom{e_{2i}}{B_{pqrs}^{(i)}+x}\right)\,  \nonumber\\
&&\cdot\frac{1}{4}\left(1-c\binom{z_1}{B_{pqrs}^{(i)}+x}\right)\,
\cdot\left(1-c\binom{z_2}{B_{pqrs}^{(i)}+x}\right)\,\nonumber\\
&&\cdot\frac{1}{2}\left(1-c\binom{\alpha }{B_{pqrs}^{(i)}+x}\right) \nonumber
\end{eqnarray}
\begin{eqnarray}
P_{pqrs}^{(i)(\bar{\psi}_{45})}&=&
\frac{1}{4}\,\left(1-c\binom{e_{2i-1}}{B_{pqrs}^{(i)}+x} \right)\,
\cdot\left(1-c\binom{e_{2i}}{B_{pqrs}^{(i)}+x}\right)\,  \nonumber\\
&&\cdot\frac{1}{4}\left(1-c\binom{z_1}{B_{pqrs}^{(i)}+x}\right)\,
\cdot\left(1-c\binom{z_2}{B_{pqrs}^{(i)}+x}\right)\, \nonumber\\
&&\cdot\frac{1}{2}\left(1 + c\binom{\alpha }{B_{pqrs}^{(i)}+x}\right) \nonumber
\end{eqnarray}
\begin{eqnarray}
P_{pqrs}^{(i)(\bar{\Phi}_{12})}&=&
\frac{1}{4}\,\left(1-c\binom{e_{2i-1}}{B_{pqrs}^{(i)}+x} \right)\,
\cdot\left(1-c\binom{e_{2i}}{B_{pqrs}^{(i)}+x}\right)\, \\
&&\cdot\frac{1}{4}\left(1+ c\binom{z_1}{B_{pqrs}^{(i)}+x}\right)\,
\cdot\left(1-c\binom{z_2}{B_{pqrs}^{(i)}+x}\right)\,\nn\\
&&\cdot\frac{1}{2}\left(1 + c\binom{\alpha }{B_{pqrs}^{(i)}+x}\right) \nonumber
\end{eqnarray}
\begin{eqnarray}
P_{pqrs}^{(i)(\bar{\Phi}_{34})}&=&
\frac{1}{4}\,\left(1-c\binom{e_{2i-1}}{B_{pqrs}^{(i)}+x} \right)\,
\cdot\left(1-c\binom{e_{2i}}{B_{pqrs}^{(i)}+x}\right)\,  \nonumber\\
&&\cdot\frac{1}{4}\left(1+ c\binom{z_1}{B_{pqrs}^{(i)}+x}\right)\,
\cdot\left(1-c\binom{z_2}{B_{pqrs}^{(i)}+x}\right)\,\nn\\
&&\cdot\frac{1}{2}\left(1 - c\binom{\alpha }{B_{pqrs}^{(i)}+x}\right) \nonumber
\end{eqnarray}
\begin{eqnarray}
P_{pqrs}^{(i)(\bar{\Phi}_{5678})}&=&
\frac{1}{4}\,\left(1-c\binom{e_{2i-1}}{B_{pqrs}^{(i)}+x} \right)\,
\cdot\left(1-c\binom{e_{2i}}{B_{pqrs}^{(i)}+x}\right)\,  \nonumber\\
&&\cdot\frac{1}{4}\left(1- c\binom{z_1}{B_{pqrs}^{(i)}+x}\right)\,
\cdot\left(1+c\binom{z_2}{B_{pqrs}^{(i)}+x}\right)\,\nn\\
&&\cdot\frac{1}{2}\left(1 - c\binom{\alpha }{B_{pqrs}^{(i)}+x}\right) \nonumber
\end{eqnarray}
The explicit expressions for the 1st  plane are the following:
\ba
\Delta_{v}^{(1)}&=&\left(
\begin{array}{cccc}
\oo{e_1}{e_3}&\oo{e_1}{e_4}&\oo{e_1}{e_5}&\oo{e_1}{e_6}\\
\oo{e_2}{e_3}&\oo{e_2}{e_4}&\oo{e_2}{e_5}&\oo{e_2}{e_6}\\
\oo{z_1}{e_3}&\oo{z_1}{e_4}&\oo{z_1}{e_5}&\oo{z_1}{e_6}\\
\oo{z_2}{e_3}&\oo{z_2}{e_4}&\oo{z_2}{e_5}&\oo{z_2}{e_6}\\
\oo{\alpha }{e_3}&\oo{\alpha }{e_4}&\oo{\alpha }{e_5}&\oo{\alpha }{e_6}
\end{array}
\right)
\nn
\ea

\ba
Y_{\bar{\psi}^{123}}^{(1)}=
\left(
\begin{array}{c}
\oo{e_1}{b_1+x}\\
\oo{e_2}{b_1+x}\\
\oo{z_1}{b_1+x}\\
\oo{z_2}{b_1+x}\\
\oo{\alpha}{b_1+x}
\end{array}
\right)
\nn
\ea

\ba
Y_{\bar{\psi}^{45}}^{(1)}=
\left(
\begin{array}{c}
\oo{e_1}{b_1+x}\\
\oo{e_2}{b_1+x}\\
\oo{z_1}{b_1+x}\\
\oo{z_2}{b_1+x}\\
1+\oo{\alpha}{b_1+x}
\end{array}
\right)
\nn
\ea

\ba
Y_{\bar{\phi}^{12}}^{(1)}=
\left(
\begin{array}{c}
\oo{e_1}{b_1+x}\\
\oo{e_2}{b_1+x}\\
1+\oo{z_1}{b_1+x}\\
\oo{z_2}{b_1+x}\\
1+\oo{\alpha}{b_1+x}
\end{array}
\right)
\ea

\ba
Y_{\bar{\phi}^{34}}^{(1)}=
\left(
\begin{array}{c}
\oo{e_1}{b_1+x}\\
\oo{e_2}{b_1+x}\\
1+\oo{z_1}{b_1+x}\\
\oo{z_2}{b_1+x}\\
\oo{\alpha}{b_1+x}
\end{array}
\right)
\nn
\ea

\ba
Y_{\bar{\phi}^{5..8}}^{(1)}=
\left(
\begin{array}{c}
\oo{e_1}{b_1+x}\\
\oo{e_2}{b_1+x}\\
\oo{z_1}{b_1+x}\\
1+\oo{z_2}{b_1+x}\\
\oo{\alpha}{b_1+x}
\end{array}
\right)
\nn
\ea

The explicit expressions for the 2nd  plane are the following:

\ba
\Delta_{v}^{(2)}&=&\left(
\begin{array}{cccc}
\oo{e_3}{e_1}&\oo{e_3}{e_2}&\oo{e_3}{e_5}&\oo{e_3}{e_6}\\
\oo{e_4}{e_1}&\oo{e_4}{e_2}&\oo{e_4}{e_5}&\oo{e_4}{e_6}\\
\oo{z_1}{e_1}&\oo{z_1}{e_2}&\oo{z_1}{e_5}&\oo{z_1}{e_6}\\
\oo{z_2}{e_1}&\oo{z_2}{e_2}&\oo{z_2}{e_5}&\oo{z_2}{e_6}\\
\oo{\alpha }{e_1}&\oo{\alpha }{e_4}&\oo{\alpha }{e_5}&\oo{\alpha }{e_6}
\end{array}
\right)
\nn
\ea

\ba
Y_{\bar{\psi}^{123}}^{(2)}=
\left(
\begin{array}{c}
\oo{e_3}{b_2+x}\\
\oo{e_4}{b_2+x}\\
\oo{z_1}{b_2+x}\\
\oo{z_2}{b_2+x}\\
\oo{\alpha}{b_2+x}
\end{array}
\right)
\nn
\ea

\ba
Y_{\bar{\psi}^{45}}^{(2)}=
\left(
\begin{array}{c}
\oo{e_3}{b_2+x}\\
\oo{e_4}{b_2+x}\\
\oo{z_1}{b_2+x}\\
\oo{z_2}{b_2+x}\\
1+\oo{\alpha}{b_2+x}
\end{array}
\right)
\nn
\ea

\ba
Y_{\bar{\phi}^{12}}^{(2)}=
\left(
\begin{array}{c}
\oo{e_3}{b_2+x}\\
\oo{e_4}{b_2+x}\\
1+\oo{z_1}{b_2+x}\\
\oo{z_2}{b_2+x}\\
1+\oo{\alpha}{b_2+x}
\end{array}
\right)
\ea

\ba
Y_{\bar{\phi}^{34}}^{(2)}=
\left(
\begin{array}{c}
\oo{e_3}{b_2+x}\\
\oo{e_4}{b_2+x}\\
1+\oo{z_1}{b_2+x}\\
\oo{z_2}{b_2+x}\\
\oo{\alpha}{b_2+x}
\end{array}
\right)
\nn
\ea

\ba
Y_{\bar{\phi}^{5..8}}^{(2)}=
\left(
\begin{array}{c}
\oo{e_3}{b_2+x}\\
\oo{e_4}{b_2+x}\\
\oo{z_1}{b_2+x}\\
1+\oo{z_2}{b_2+x}\\
\oo{\alpha}{b_21+x}
\end{array}
\right)
\nn
\ea

The explicit expressions for the 3rd plane are the following:

\ba
\Delta_{v}^{(2)}&=&\left(
\begin{array}{cccc}
\oo{e_5}{e_1}&\oo{e_5}{e_2}&\oo{e_5}{e_3}&\oo{e_5}{e_4}\\
\oo{e_6}{e_1}&\oo{e_6}{e_2}&\oo{e_6}{e_3}&\oo{e_6}{e_4}\\
\oo{z_1}{e_1}&\oo{z_1}{e_2}&\oo{z_1}{e_3}&\oo{z_1}{e_4}\\
\oo{z_2}{e_1}&\oo{z_2}{e_2}&\oo{z_2}{e_3}&\oo{z_2}{e_4}\\
\oo{\alpha }{e_1}&\oo{\alpha }{e_4}&\oo{\alpha }{e_5}&\oo{\alpha }{e_6}
\end{array}
\right)
\nn
\ea

\ba
Y_{\bar{\psi}^{123}}^{(3)}=
\left(
\begin{array}{c}
\oo{e_5}{b_3+x}\\
\oo{e_6}{b_3+x}\\
\oo{z_1}{b_3+x}\\
\oo{z_2}{b_3+x}\\
\oo{\alpha}{b_3+x}
\end{array}
\right)
\nn
\ea

\ba
Y_{\bar{\psi}^{45}}^{(3)}=
\left(
\begin{array}{c}
\oo{e_5}{b_3+x}\\
\oo{e_6}{b_3+x}\\
\oo{z_1}{b_3+x}\\
\oo{z_2}{b_3+x}\\
1+\oo{\alpha}{b_3+x}
\end{array}
\right)
\nn
\ea

\ba
Y_{\bar{\phi}^{12}}^{(3)}=
\left(
\begin{array}{c}
\oo{e_5}{b_3+x}\\
\oo{e_6}{b_3+x}\\
1+\oo{z_1}{b_3+x}\\
\oo{z_2}{b_3+x}\\
1+\oo{\alpha}{b_3+x}
\end{array}
\right)
\ea

\ba
Y_{\bar{\phi}^{34}}^{(3)}=
\left(
\begin{array}{c}
\oo{e_5}{b_3+x}\\
\oo{e_6}{b_3+x}\\
1+\oo{z_1}{b_3+x}\\
\oo{z_2}{b_3+x}\\
\oo{\alpha}{b_3+x}
\end{array}
\right)
\nn
\ea

\ba
Y_{\bar{\phi}^{5..8}}^{(3)}=
\left(
\begin{array}{c}
\oo{e_5}{b_3+x}\\
\oo{e_6}{b_3+x}\\
\oo{z_1}{b_3+x}\\
1+\oo{z_2}{b_3+x}\\
\oo{\alpha}{b_3+x}
\end{array}
\right)
\nn
\ea



\bigskip
\medskip

\bibliographystyle{unsrt}

\end{document}